\documentclass[traditabstract]{aa}

\usepackage{graphicx}
\usepackage{txfonts}
\usepackage{array}
\usepackage{amssymb}
\usepackage{natbib}
\usepackage{lscape}
\usepackage{rotating}
\usepackage{longtable}
\newcommand{\duz}{Schulte\,12}
\newcommand{\kms}{km\,s$^{-1}$}
\newcommand{\xmm}{{\sc{XMM}}\emph{-Newton}}
\newcommand{\ch}{\emph{Chandra}}
\newcommand{\sw}{{\em Swift}}
\usepackage{color}

\begin{document}

\title{Variations on a theme -- the puzzling behaviour of \duz\thanks{Based on X-ray observations collected with {\it ASCA, Suzaku, Chandra} and especially {\it{Swift}} and \xmm , an ESA Science Mission with instruments and contributions directly funded by ESA Member States and the USA (NASA). Optical photometry from {\it NSVS, Integral-OMC,} and {\it ASAS-SN} are also used, as well as optical spectroscopy from the Hermes and Carmenes instruments. Hermes is installed on the Mercator Telescope, which is operated on the island of La Palma by the Flemish Community, at the Spanish Observatorio del Roque de los Muchachos of the Instituto de Astrof\'{\i}sica de Canarias.}}

\author{Ya\"el~Naz\'e\inst{1}\thanks{F.R.S.-FNRS Research Associate.}
\and Gregor~Rauw\inst{1} \and Stefan Czesla\inst{2} \and Laurent Mahy\inst{3} \and Fran Campos\inst{4}
}

\institute{Groupe d'Astrophysique des Hautes Energies, STAR, Universit\'e de Li\`ege, Quartier Agora (B5c, Institut d'Astrophysique et de G\'eophysique), All\'ee du 6 Ao\^ut 19c, B-4000 Sart Tilman, Li\`ege, Belgium\\
  \email{ynaze@uliege.be}
  \and Hamburger Sternwarte, Universit\"at Hamburg, Gojenbergsweg 112, 21029, Hamburg, Germany
  \and Instituut voor Sterrenkunde, KU Leuven, Celestijnlaan 200D, Postbus 2401, 3001, Leuven, Belgium
  \and Observatori Puig d’Agulles, Passatge Bosc 1, 08759 Vallirana, Barcelona, Spain
}

\authorrunning{Naz\'e et al.}
\titlerunning{\duz }
\abstract{One of the first massive stars detected in X-rays, \duz\ has remained a puzzle in several aspects. In particular, its extreme brightness both in the visible and X-ray ranges is intriguing. Thanks to \sw\ and \xmm\ observations covering $\sim$5000\,d, we report the discovery of a regular 108\,d modulation in X-ray flux of unknown origin. The minimum in the high-energy flux appears due to a combination of increased absorption and decreased intrinsic emission. We examined in parallel the data from a dedicated spectroscopic and photometric monitoring in the visible and near-IR domains, complemented by archives. While a similar variation timescale is found in those data, they do not exhibit the strict regular clock found at high energies. Changes in line profiles cannot be related to binarity but rather correspond to non-radial pulsations. Considering the substantial revision of the distance of \duz\ from the  second {\it GAIA} data release, the presence of such oscillations agrees well with the evolutionary status of \duz, as it lies in an instability region of the HR diagram.  }
\keywords{stars: early-type -- stars: massive -- stars: winds -- X-rays: stars -- supergiants -- stars: variable: general -- stars: individual: \object{\duz}}
\maketitle

\section{Introduction}
In 1978, the $Einstein$ observatory detected for the first time X-rays associated to massive stars \citep{har79,ku79, sew79}. Among these was Cyg\,OB2\,\#12 or \duz\ ($V$=12.5), a highly reddened and peculiar star known for several decades \citep{mor54}. \duz\  was classified as an early B hypergiant of spectral type B3--4Ia+ \citep{cla12}. With its extreme brightness, the star would be placed in the Hertzprung-Russell diagram above the Humphreys-Davidson limit. This explains why \duz\ was often considered as a luminous blue variable (LBV) candidate \citep{cla05}. In support of this classification, evidence for changes in spectral type (B3--B8) have been presented by \citet{mas91} and \citet{kim07}. Variations of the temperature were later refuted by \citet{cla12} because of the low signal-to-noise ratio of the data, prohibiting changes to reach a low significance level, but also because changes in apparent spectral type may not necessarily reflect a varying photosphere hence could agree with a constant temperature. A lack of helium enrichment is also at odds with an evolved star status \citep{cla12}. Furthermore, no surrounding ejected nebulosities seem to exist \citep{kob12,osk17}. 

The early X-ray detection poses a puzzle which could not be solved in the four decades elapsed since then. Indeed, the X-ray emission of \duz\ is bright ($\log[L_{\rm X}/L_{\rm BOL}]=-6.1$) and presents a hard ($kT\sim 2$\,keV) component \citep{rau11}. This is clearly at odds with the properties of LBVs, which generally present very faint (undetected) X-ray emissions \citep{naz12lbv}. Besides, it is also at odds with typical properties of massive stars. The intrinsic X-rays from massive stars with fast winds (thousands of \kms) are generally thought to come from embedded wind shocks \citep[e.g.][]{fel97}, which generate soft ($kT=0.3-0.6$\,keV) and modestly bright ($\log[L_{\rm X}/L_{\rm BOL}]\sim-7$ for O-stars) X-ray emissions \citep{osk05,naz09,rau15}. In this context, it is interesting to note that \duz\ actually has a very slow wind ($v_{\infty}=400$\,\kms, \citealt{cla12} or even 150\,\kms, \citealt{klo04}), incompatible with high-temperature plasma - though larger velocities have also been proposed; see \citet{lei82}. 

There is however another Galactic LBV that displays bright and hard X-rays: $\eta$\,Carinae. In this case, they arise from the face-on collision between the two winds of the components in this massive binary \citep[see e.g.][and references therein]{ham14}. Similarly, the LBV HD\,5980 in the Small Magellanic Cloud also emits bright and hard X-rays linked to a wind--wind collision \citep[and references therein]{naz18}. A telltale signature of such colliding winds is variability \citep[for a review, see][]{rau16}. Indeed, the intrinsic strength of the collision changes with the stellar separation in eccentric binaries and the absorption may also vary as the shocked plasma is alternately seen through the two different winds \citep{ste92}. Because they are linked to the orbital configuration, such changes recur with the orbital period. For \duz, variations of its X-ray emission have been reported several times \citep{rau11,yos11,caz14} but no periodicity was found. In view of the apparently monotonic decrease in X-ray fluxes from 2004 to 2013, an alternative scenario where X-rays are produced by fading shocks between slow-moving ejecta and fast winds after an eruption could not be rejected \citep{caz14}.

The possible presence of a companion to \duz\ has been considered in the past \citep{sou80} and has subsequently been revived in a few recent studies. First, \citet{cab14} reported the detection of a close visual companion at 64\,mas with $\Delta V=2.3$. \citet{mar16a} confirmed its presence and constrained its orbital period to 100--200\,yrs from its slight change of position. These latter authors also found a third object further out at 1.2\arcsec with $\Delta m=4.8$. However, such distant companions are unlikely to produce an X-ray bright wind--wind collision as plasma density would be too small at such large separations. Until now, X-ray bright wind--wind collisions have been found in massive binaries with periods of a few years (up to a decade in 9\,Sgr; \citealt{rau16b}), not decades or centuries. Secondly, small line-profile and radial-velocity (RV) changes were reported by \citet{klo04} and \citet{che13} but they could not confirm or rule out the binarity hypothesis \citep{che13}. Only a double-lined spectroscopic binary seemed excluded by the data \citep[see also][]{cla12}. Finally, \citet{osk17} studied a high-resolution X-ray spectrum of \duz\ and found rather strong $f$ lines in $fir$ triplets from He-like ions. While the Si\,{\sc xiii} triplet is compatible with the absence of depopulation of the upper level of the $f$ line, the Mg\,{\sc xi} lines suggest some depopulation, though not due to UV radiation but to collisions: there would thus be dense plasma where X-rays arise. This could be compatible with X-ray generation very close to the photosphere or in a wind--wind collision. The latter scenario is further supported by the (broad and unshifted) profiles of X-ray lines and the lack of very strong absorption. \citet{osk17} further suggested a late-O type for the optically closest companion, if there was a wind--wind interaction between this star and the LBV (which remains to be demonstrated). However, they interpreted the observed X-ray variability as being random and incompatible with a companion on a long (decades) orbit.

While important information has been gained in recent years, the nature of \duz\ is still not fully understood. Time is thus ripe to perform an in-depth study, with a thorough X-ray and optical monitoring (described in Section 2). The derived results led to several surprises (Sections 3 and 4 for X-ray and optical/near-IR findings, respectively). Their possible interpretation is discussed in Section 5 and then summarized in Section 6.

\section{Observations and data reduction} 

\subsection{Optical and near-IR domains}

\subsubsection{Photometry}

Photometry of \duz\ was obtained in both the $Ic$ and $V$ filters at the private observatory of one of the current authors (F.C.), situated in Vallirana (near Barcelona, Spain) and equipped with a Newton telescope of 20cm diameter (with f/4.7 and a German equatorial mount). The camera is a CCD SBIG ST-8XME (KAF 1603ME). The exposures were typically of 180\,s and 120\,s duration for the $V$ and $Ic$ filters, respectively. 

The images were corrected for bias, dark current, and flat-field in the usual way using the data-reduction software Maxim DL v5\footnote{https://diffractionlimited.com/help/maximdl/MaxIm-DL.htm}. The photometry was extracted with the FotoDif v3.93 software, using as comparison star SAO 49783 (TYC 3157-195-1). Four stars (TYC 3157-1310-1, TYC 3157-603-1, TYC 3161-1269-1, TYC 3157-463-1) were further measured to check the stability of the solution. 

\begin{table}
\centering
\caption{Journal of the optical and near-IR spectroscopic observations. HJD correspond to dates at mid-exposure and S/Ns are evaluated at 7525\AA.}
\label{jouropt}
\setlength{\tabcolsep}{3.3pt}
\begin{tabular}{cccccc}
\hline\hline
\multicolumn{3}{c}{Hermes} & \multicolumn{3}{c}{Carmenes} \\
Date & HJD & $SNR$ & Date & HJD & $SNR$ \\
YYMMDD & --2\,450\,000 & & YYMMDD & --2\,450\,000 & \\
\hline
20100626 & 5373.640 & 240 & 20170524 & 7898.611 & 250 \\
20110709 & 5751.665 & 260 & 20170607 & 7912.647 & 310 \\
20110828 & 5801.572 & 170 & 20170623 & 7928.628 & 290 \\
20110903 & 5807.605 & 220 & 20180502 & 8241.666 & 160 \\
20121115 & 6247.331 & 210 & 20180512 & 8251.647 & 50  \\
20131106 & 6603.394 & 200 & 20180525 & 8264.632 & 180 \\
20140606 & 6814.652 & 220 & 20180530 & 8269.608 & 250 \\
20150710 & 7213.631 & 220 & 20180614 & 8284.568 & 290 \\
20151128 & 7355.350 & 60  & 20180620 & 8290.648 & 250 \\
20160623 & 7562.663 & 170 & 20180625 & 8295.573 & 280 \\
20160708 & 7578.467 & 140 & 20180630 & 8300.544 & 200 \\
20170529 & 7902.611 & 240 & 20180714 & 8314.553 & 80  \\
         &          &     & 20180721 & 8321.585 & 180 \\
         &          &     & 20180804 & 8335.484 & 250 \\
         &          &     & 20180811 & 8342.572 & 250 \\
         &          &     & 20180817 & 8348.455 & 310 \\
         &          &     & 20180825 & 8356.455 & 240 \\
         &          &     & 20180904 & 8366.396 & 210 \\
         &          &     & 20180913 & 8375.367 & 110 \\
\hline      
\end{tabular}
\end{table}

\subsubsection{Spectroscopy}
\duz\ was observed 12 times between 2010 and 2017 with the Hermes spectrograph \citep{ras11} installed on the 1.2\,m Mercator telescope on La Palma (Spain). Given the magnitude of the star, between 2 and 7 consecutive exposures of 1800\,s (depending on the weather conditions) were summed up to increase S/N. The data were taken in the high-resolution fibre mode, which has a resolving power of $R \sim 85000$. The spectra cover the 4000--9000\AA\ wavelength domain but because of the high extinction of the star, the range below 5750\AA\ was not used for the present analysis. The raw exposures were reduced using the dedicated Hermes pipeline and we worked with the extracted cosmic-removed, merged spectra afterwards.

A set of 19 observations were obtained with the Carmenes \citep[Calar Alto high-Resolution search for M dwarfs with Exoearths with near-IR and optical Echelle Spectrographs;][]{CARMENES} instrument at the 3.5\,m telescope of the Centro Astron\'omico Hispano Alem\'an at Calar Alto Observatory (Spain). These observations were granted via the OPTICON common time allocation process (program IDs 2018A/004 and 2018B/001) and through German open time (program ID F17-3.5-002). Carmenes features a dichroic that shares the light between two separate echelle spectrographs: the first one covering the optical domain from about 5200 to 9600\,\AA\ at a spectral resolving power near $R = 94\,600$, and the second covering the near-IR from about 9600\,\AA\ to 1.71 $\mu$m at $R = 80\,400$. All data were obtained in service mode at a rate of about one observation every 10\,d during our main observing campaign in May--August 2018, and the exposure times were $4 \times 5$\,min per observing night. The data were reduced using the Carmenes pipeline \citep{cab16}. The Carmenes spectra taken the same night were combined to improve S/Ns.

Table \ref{jouropt} lists the dates of these spectra, as well as their S/Ns around 7525\AA. Because of the strong extinction, typical S/Ns  vary from $\sim$50 at 5800\AA\ to $\sim$300 in the near-IR. To obtain the best diagnostics, we focused on wavelength intervals with strong stellar lines and limited contamination by interstellar or telluric lines. However, small contamination could not always be totally avoided. We therefore performed a correction for telluric lines around H$\alpha$, He\,{\sc i}\,$\lambda$\,7065\AA, and He\,{\sc i}\,$\lambda$\,10830\AA, as well as Paschen lines at 8500--8900\AA\ and 1$\mu$m. This was done within IRAF using the telluric template of \citet{hin00} in the optical range and a telluric spectrum calculated for Calar Alto with {\sc molecfit} \citep{sme15} in the near-IR. As a last step, the spectra were normalized over limited wavelength windows using splines of low order. 

\subsection{X-ray domain}

\subsubsection{\xmm }
The Li\`ege group have obtained ten observations of the Cyg\,OB2 region since the launch of \xmm. The oldest ones were reported in \citet{deb06}, \citet{rau11}, \citet{naz12}, and \citet{caz14}. \duz\ also appears in six additional \xmm\ exposures centred on PSR J2032+4127. However, in these exposures it is far off-axis and only in the field-of-view of the MOS2 detectors. Moreover, stray light from Cyg\,X-3 contaminates the source (further discussed below). All datasets were reduced with SAS v16.0.0 using calibration files available in Fall 2017 and following the recommendations of the \xmm\ team\footnote{SAS threads, see \\ http://xmm.esac.esa.int/sas/current/documentation/threads/ }. No barycentric correction was applied to event arrival times.

The EPIC observations were taken in the full-frame mode and with the medium filter (to reject optical/UV light), except for ObsID 0677980601 where the large window mode was used for the MOS cameras to avoid pile-up in Cyg\,OB2\,\#9 (which was then at its maximum). After pipeline processing, the data were filtered to keep only best-quality data ({\sc{pattern}} of 0--12 for MOS and 0--4 for pn). Background flares were detected in several observations (Revs 0896, 0911, 1353, 1355, 2114, 3097). Only times with a count rate above 10.\,keV lower than 0.2--0.3\,cts\,s$^{-1}$ (for MOS) or 0.4\,cts\,s$^{-1}$ (for pn) were kept. A source detection was performed on each EPIC dataset using the task {\it edetect\_chain} on the 0.3--2.0\,keV (soft) and 2.0--10.0\,keV (hard) energy bands and for a minimum log-likelihood of 10. This task searches for sources using a sliding box and determines the final source parameters from point spread function (PSF) fitting; the final count rates correspond to equivalent on-axis, full-PSF count rates (Table~\ref{journal}). It may be noted in this table that the count rates of \duz\ in the off-axis observations are unreliable, as demonstrated by their much larger hardness ratios (due to uncorrected contamination by stray light).

We then extracted EPIC spectra of \duz\ using the task {\it{especget}} in circular regions of 35\arcsec\ radius (to avoid nearby sources) centred on the Hipparcos positions (reported in Simbad). For ObsID 0793183001, the source appears so far off-axis that a larger ellipse was needed to extract it. On the contrary, in a few other cases the extraction radius had to be reduced for one of the EPIC cameras to avoid nearby CCD gaps. Whenever the core of the  source fell in one such gap, the spectrum was not extracted to avoid large uncertainties. For the background, a circular region of 35\arcsec\ radius was chosen in a region devoid of sources and as close as possible to the target. Dedicated ARF and RMF response files, which are used to calibrate the flux and energy axes, respectively, were also calculated by this task. EPIC spectra were grouped, with {\it{specgroup}}, to obtain an oversampling factor of five and to ensure that a minimum S/N of 3 (i.e.\ a minimum of 10 counts) was reached in each spectral bin of the background-corrected spectra. 

\subsubsection{\sw}
At our request, the Cyg\,OB2 region was observed 18 times by the Neil Gehrels \sw\ Observatory. In addition, the same area was the subject of a short campaign in June--July 2015 and it serendipitously appears from time to time in the field of view of observations dedicated to PSR J2032+4127 \citep{pet18}, 3EGJ2033+4118, or WR\,144. All these XRT data taken in Photon Counting mode were retrieved from the HEASARC archive centre and were further processed locally using the XRT pipeline of HEASOFT v6.22.1 with calibrations v20170501. Because of the high extinction towards Cyg\,OB2, no optical loading is expected for XRT data of \duz. The source spectra were extracted within Xselect using a circular region of 35\arcsec\ radius; a nearby region devoid of any source was used for the background. The spectra were binned using {\it grppha} in a similar manner as the \xmm\ spectra. The adequate RMF matrix from the calibration database was used whilst specific ARF response matrices were calculated for each dataset using {\it xrtmkarf}, considering the associated exposure map. Corrected count rates in the same energy bands as \xmm\ were obtained for each observation from the UK online tool\footnote{http://www.swift.ac.uk/user\_objects/}, and they are reported in Table~\ref{journal}.

$UVOT$ photometry was taken in parallel with the X-ray data using a variety of filters. Following the recommendations of the \sw\ team\footnote{http://www.swift.ac.uk/analysis/uvot/mag.php}, magnitudes were extracted from Level2 files using an extraction region of 5\arcsec\ radius for the source (and the same coordinates as in X-rays) while a nearby region of 15\arcsec\ radius and devoid of sources was used as background. The resulting photometry is provided in the last column of Table \ref{journal}, where the reported errors correspond to the combination of systematic and statistical errors. No significant correlation is found between X-ray count rates and UV photometry. 

\subsubsection{\ch }
The massive stars of Cyg\,OB2 are X-ray bright, and therefore suffer from pile-up in usual \ch -ACIS observations. However, \duz\ was also observed twice with \ch -HETG (Table~\ref{journal}). We reprocessed these data using CIAO 4.9 and CALDB 4.7.3. In the first dataset (ObsID=2572), \duz\ appears far off-axis: its high-resolution spectrum is of low quality \citep{osk17}, but this position, leading to a larger PSF, allows its zero-order spectrum to avoid pile-up. This zero-order spectrum was therefore extracted using {\it specextract} in a circle of radius 20\,px (9.8\arcsec) centred on the Simbad coordinates of the star and a nearby circle of 14.8\arcsec\ radius was used as background region. The second exposure (ObsID=16659) was pointed at \duz, and therefore the high-resolution HEG and MEG spectra were extracted, combining the orders +1 and --1 for each grating, in addition to the zeroth-order spectrum. Specific detector response matrices (RMF and ARF) were calculated to obtain energy- and flux-calibrated spectra and a grouping similar to that used for \xmm\ spectra was applied.

\subsubsection{JAXA facilities}
{\it ASCA} observed Cyg\,OB2 in April 1993 while {\it Suzaku} did so in December 2007. We downloaded these archival data from the HEASARC website and reduced them according to recommendations\footnote{See cookbooks on https://heasarc.gsfc.nasa.gov/docs/asca/abc/abc.html and http://heasarc.gsfc.nasa.gov/docs/suzaku/analysis/abc/ }.

The {\it ASCA} observation of Cyg\,OB2 consists of several datasets because of two different data modes (BRIGHT and BRIGHT2) and bit rates (high and medium). In all of them, \duz\ appears in the middle of stray light due to Cyg\,X-3. We extracted in Xselect the individual SIS-0 and SIS-1 spectra using a circle of 80\arcsec\ radius, to avoid the emissions of the neighbouring stars, and considering the background from the surrounding annulus (with an external radius of 199\arcsec) to best compensate the stray light contamination. Data from high and medium bit rates were combined, but those from the two modes and two detectors were extracted separately, leading to four spectra in total. Following recommendations, RMF and ARF responses were calculated using the task {\sc sisrmg} (for grades 0234) and the task {\sc ascaarf} (for a point source and simple case), respectively. A grouping similar to that used for \xmm\ spectra was applied with {\sc grppha}. However, the final spectra were found to be extremely noisy and therefore unusable, so they were discarded from our analysis.

In {\it Suzaku} observations, we extracted in Xselect the individual XIS-0, XIS-1, and XIS-3 spectra of \duz\ in a circle of 80\arcsec\ radius, to avoid the neighbouring emissions, and we used a nearby circle of 100\arcsec\ radius for the background. Response files were automatically calculated in Xselect using the calibration database of June 2016. A grouping similar to that used for \xmm\ spectra was applied with {\sc grppha}.

\begin{figure}
  \begin{center}
\includegraphics[width=8.5cm]{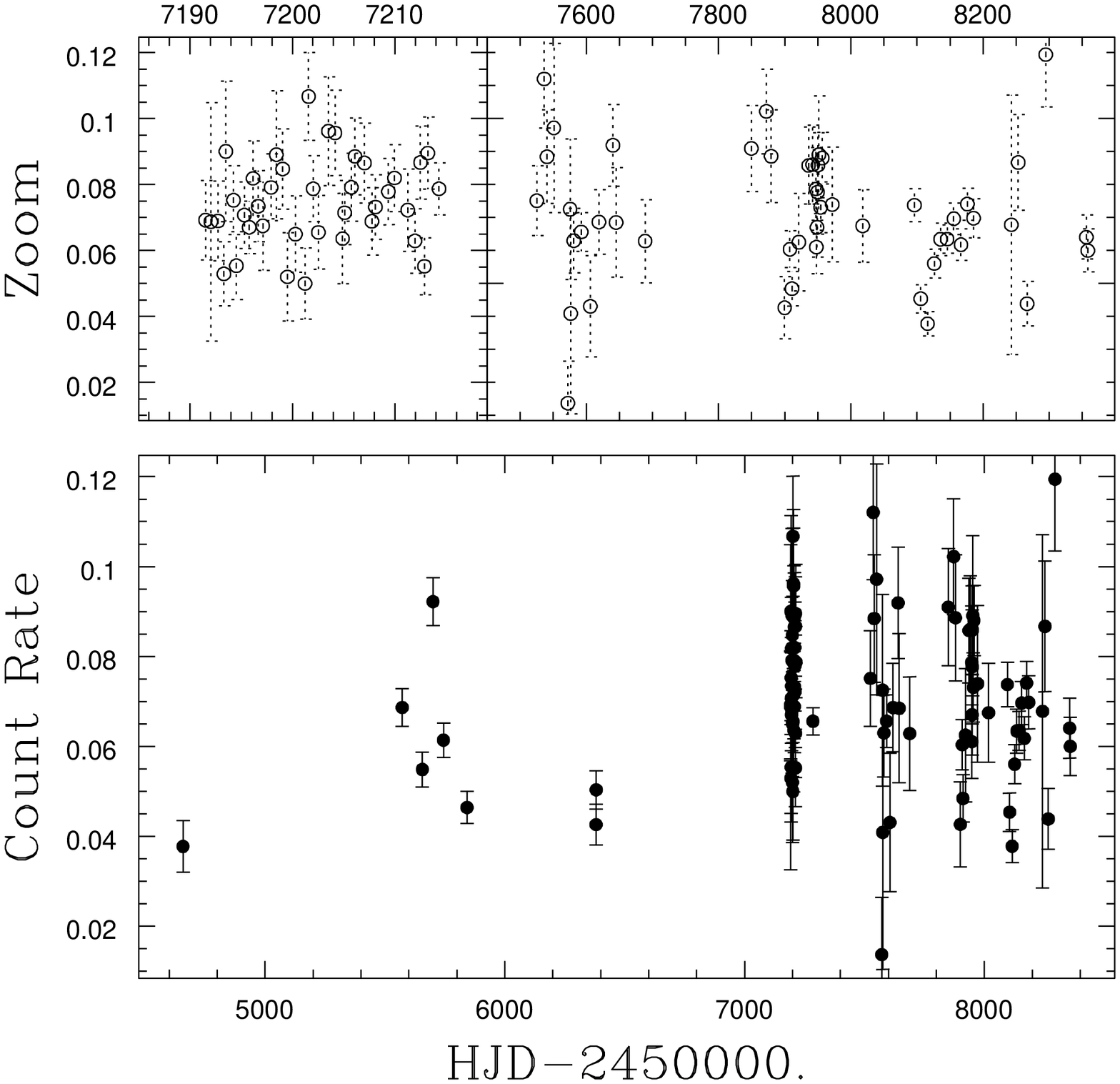}
  \end{center}
\caption{\sw\ light curve of \duz.}
\label{lc}
\end{figure}

\begin{figure*}
  \begin{center}
\includegraphics[width=8.5cm]{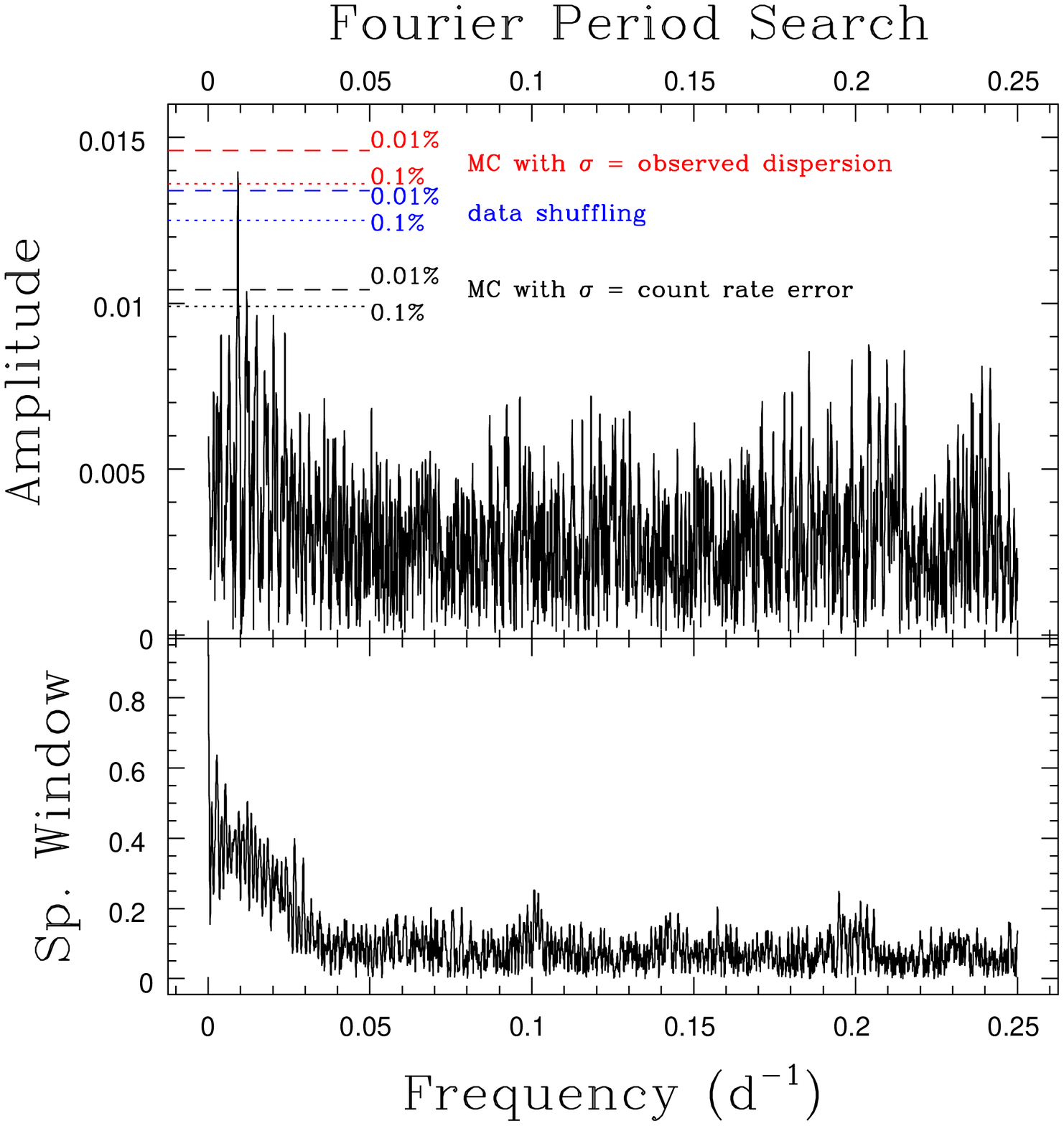}
\includegraphics[width=8.5cm]{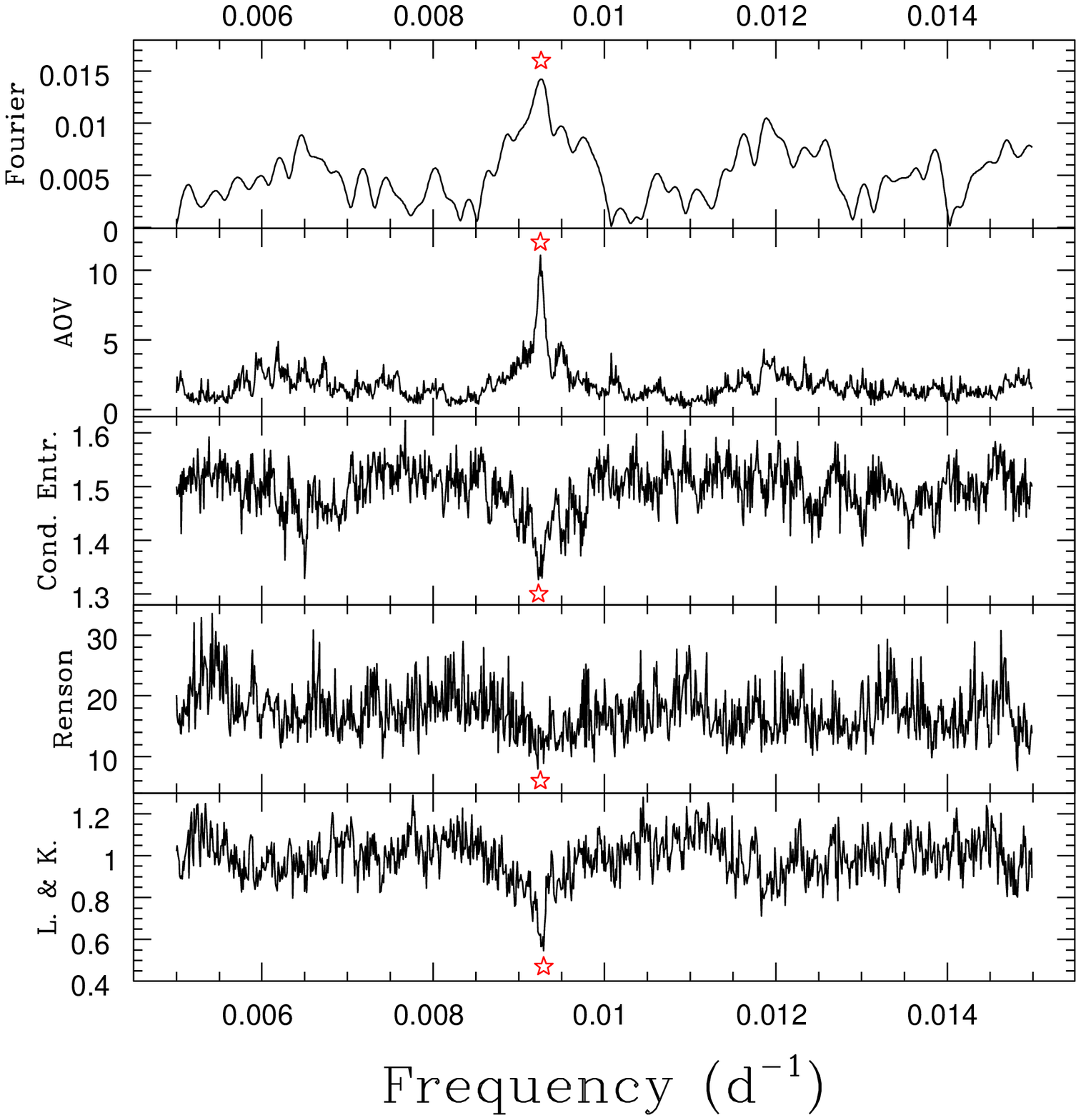}
\includegraphics[width=8.5cm]{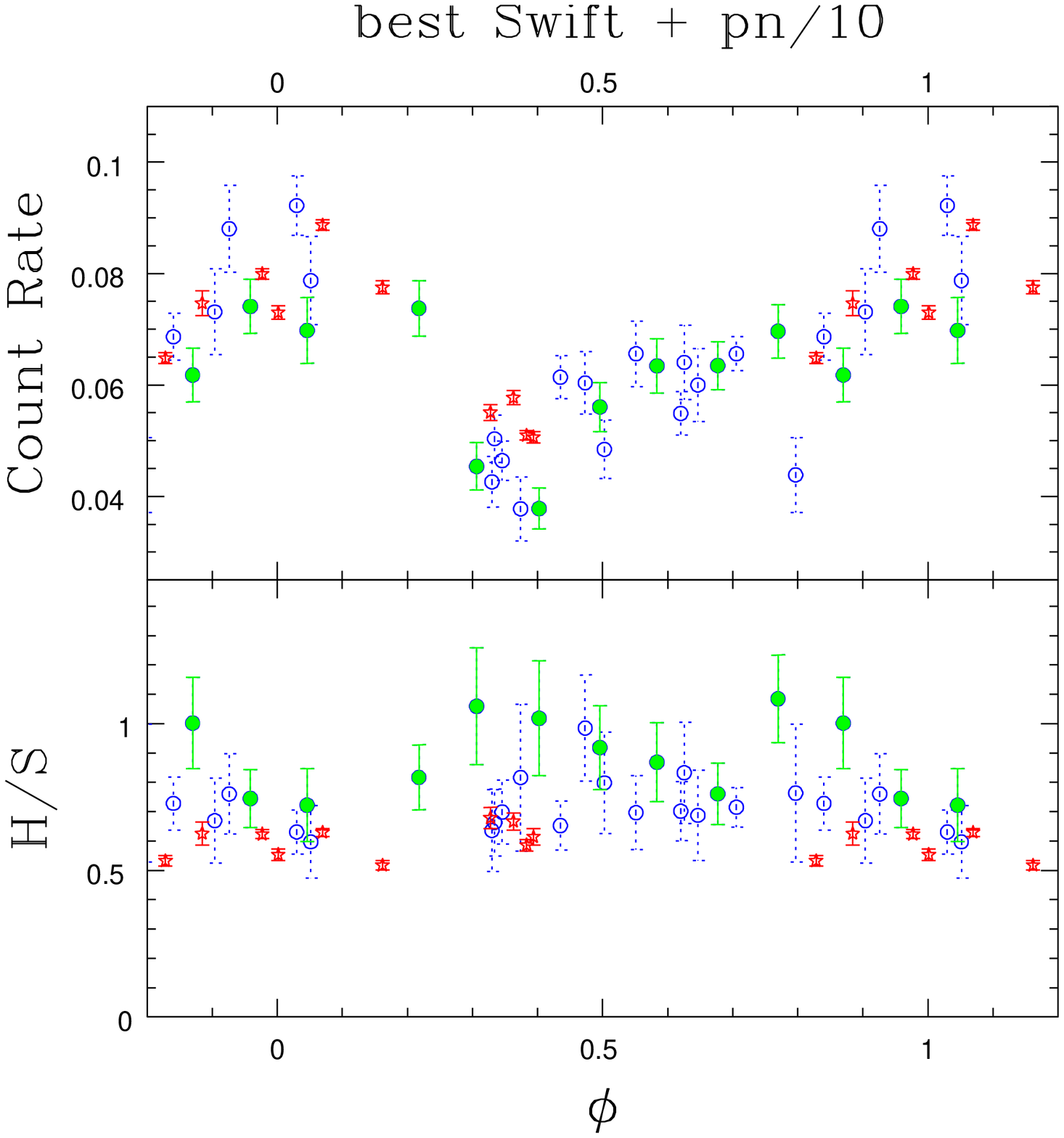}
\includegraphics[width=8.5cm]{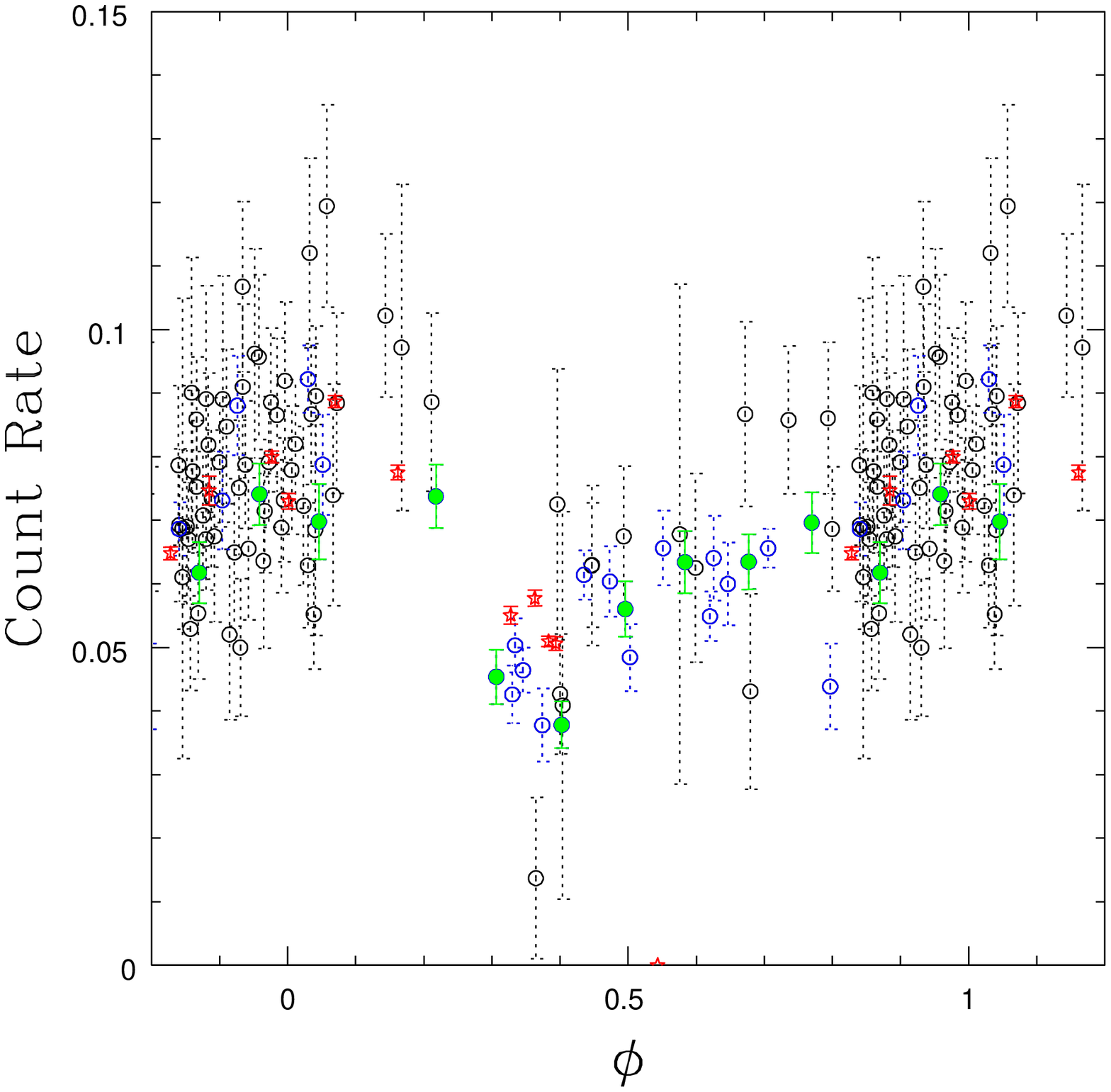}
  \end{center}
\caption{{\it Top left:} Fourier periodogram calculated from the \sw\ light curve, showing a clear peak at low frequencies. Significance levels of 0.1\% and 0.01\%, derived from tests (see text for details), are shown as dotted and dashed lines, respectively. {\it Top right:} Comparison of period search results on the light curve composed of \sw\ data and scaled EPIC-pn data with different methods: the same peak (maxima for Fourier and AOV, minima for Renson, Lafler \& Kinman, and conditional entropy) is detected in all cases and marked by a star. {\it Bottom left:} Phase-folded X-ray light curve, using $P=107.99$\,d and $T_0$=2\,458\,288.654. Our 2018 \sw\ monitoring campaign of \duz\ is shown in green and other high-quality (error on count rate $< 0.008$\,cts\,s$^{-1}$) \sw\ points in blue; EPIC-pn count rates divided by ten are shown as red stars; we note the good agreement between all data. {\it Bottom right:} Same as previous, but with the lower-quality \sw\ data shown in dotted black. }
\label{lc2}
\end{figure*}

\section{X-ray results}

\subsection{Light curves}

As found previously with \xmm\ \citep{caz14}, the \sw\ light curve in the total band clearly displays variations of a large amplitude, with rates appearing mostly between 0.04 and 0.10\,cts\,s$^{-1}$ (Fig. \ref{lc}). These changes do not appear correlated to hardness changes. To better characterise these variations in X-ray brightness, we applied a set of period-search algorithms: (1) the Fourier algorithm adapted to sparse/uneven datasets \citep[see also its recent rediscovery by \citealt{zec09} - these papers also note that the method of \citealt{sca82}, while popular, is not fully statistically correct]{hmm,gos01}, (2) three binned analyses of variances (\citealt{whi44}, \citealt{jur71} which is identical, with no bin overlap, to the ``pdm'' method of \citealt{ste78}, and \citealt{cuy87} - which is identical to the ``AOV'' method of \citealt{sch89}), (3) conditional entropy \citep[see also \citealt{gra13}]{cin99,cin99b}, and (4) two different string length methods \citep{lafkin,renson}. The Fourier periodogram (top left panel of Fig. \ref{lc2}) clearly reveals a peak at $P=108.11\pm0.32$\,d - the error here corresponds to the accuracy with which the period can be determined (i.e. a small fraction of the peak width). As expected from the spectral window (bottom part of the top left panel of Fig. \ref{lc2}), the peak is accompanied by secondary peaks, which are aliases of the main one and are not discussed further. The periodicity is confirmed by the other methods (that find $P$ in the range 107.99--109.05\,d), even though some of them, especially the string length methods, are less sensitive. If we divide the \xmm\ EPIC-pn count rates by a factor of ten and make a period search on all \xmm\ and \sw\ data combined, then a period of $P=107.99\pm0.23$\,d is found by the Fourier method, with $P=107.64-108.45$\,d for the others (top right panel of Fig. \ref{lc2}).

The significance of this X-ray period can be readily estimated for the AOV method using the $F$ statistics. The amplitude in the AOV periodogram corresponds to the ratio of the variance of the individual means in M phase bins and the sum over all phase bins of the variances about the mean within each bin. Amplitude therefore follows an F-distribution with $(M-1, N-M)$ degrees of freedom, with $M$ the number of bins (chosen equal to 10) and $N$ the number of data points (e.g. $N$=96 for \sw\ data and 106 for \sw+\xmm\ data). For the top-right AOV panel of Fig. \ref{lc2}, significance levels of $<$0.01\% are formally reached for amplitudes above 4.3.: the observed peak amplitude therefore corresponds to a significance level well below 0.01\%.

For the Fourier method, no analytical formula can be used to determine the significance level of the highest peak in case of uneven sampling, but the significance level can be estimated in other ways. First, we performed two sets of 10\,000 Monte-Carlo simulations considering the observation times and drawing at random the individual count rates from a Gaussian distribution with the same mean for all simulations and a standard deviation equal to (1) the observational error on the count rate at each time or (2) the dispersion of the observed count rates. The first test probes the existence of the periodicity against a constant signal considering only the observational errors, while the second one tries to reproduce the variability range of the data assuming random variations, thereby testing a different null hypothesis. Secondly, we performed another test by shuffling the data 10\,000 times: the count rate values were randomly associated with the observing times, again testing the periodicity considering the full range of the count rates. The 0.1\% and 0.01\% significance levels derived from these tests are shown on the top left panel of Fig. \ref{lc2}. With an amplitude of 0.014\,cts\,s$^{-1}$, the observed peak is associated to a significance level below 0.01\% for the first MC test and the shuffling test, and of 0.02\% for the second MC simulation. With such low significance levels, the detection of the X-ray periodicity is thus undoubtly significant.

The phase-folded light curve (bottom panels of Fig. \ref{lc2}) further underlines the good agreement between \sw\ and \xmm\ data. The confirmation of the period,  in independent instruments and for data taken years apart, shows that the X-ray periodicity exists and is stable at least over the $\sim$5000\,d interval covered by \sw\ and \xmm. 

\subsection{Spectral fits}

We fitted the spectra (Fig. \ref{specfig}) under Xspec v12.9.1p, considering reference solar abundances from \citet{asp09}. As X-ray lines are clearly detected on both low- and high-resolution spectra \citep[see also][]{osk17}, absorbed optically thin thermal-emission models of the type $tbabs\times phabs \times \sum apec$ were considered, as is usual for massive stars. The first absorption component represents the interstellar contribution, fixed to $2.04\times10^{22}$\,cm$^{-2}$, a value calculated using the reddening of the target ($E(B-V)=3.33$, \citealt{whi}) and the conversion factor from \citet{gud12}. 

\begin{figure}
  \begin{center}
\includegraphics[width=8.5cm]{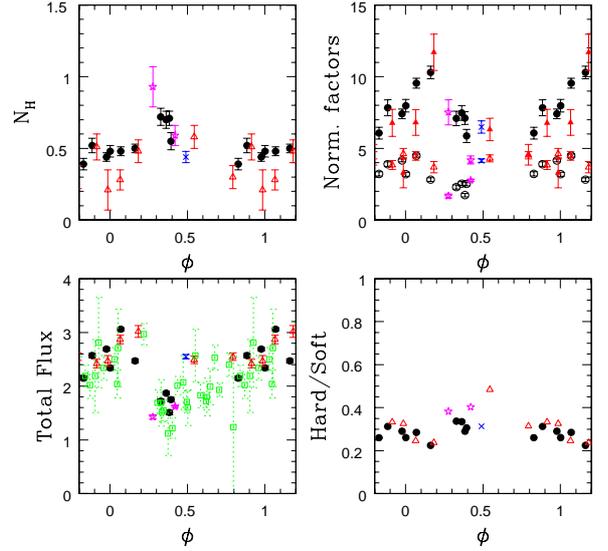}
  \end{center}
\caption{Results of the spectral fits: circumstellar absorption (top left), normalization factors (top right - the $norm_1$ (resp. $norm_2$) values appear in the upper (resp. lower) part, with filled (resp. empty) circles for \xmm\ data), observed flux (bottom left), and ratio of unabsorbed hard and soft fluxes (bottom right). The results for \xmm\ data are shown as black circles except for the stray light-contaminated cases displayed as red triangles; magenta stars indicate \ch\ data; blue crosses the {\it Suzaku} observation; and \sw\ data are shown as empty green squares (shown for fluxes only). The phase folding is done as in Fig. \ref{lc2}.}
\label{figfit}
\end{figure}

We first fitted the \xmm\ EPIC spectra, considering all available MOS and pn data simultaneously. Two thermal components were sufficient to achieve a good fit. Several combinations of absorption and temperature however yielded fits of similar quality (i.e. similar $\chi^2$), again a usual feature of the spectra of massive stars. We decided to fix the two temperatures to 0.75 and 1.9\,keV, the most common values, leaving the absorption degeneracy, and repeated the fits. As we suspected, the fit quality was similar to what we had before. The same model was used for fitting \ch\ and {\it Suzaku} spectra as well as the \sw\ spectra with the highest S/Ns (error on count rate below 0.08\,cts\,s$^{-1}$, i.e. data shown as green and blue points in Fig. \ref{lc2}). The results of these fits are provided in Table \ref{fits} and Fig. \ref{figfit}. The fitted fluxes (and the normalization factors) closely follow the count rates, as expected (see Fig. \ref{lc2}). The hardness ratios do not vary significantly, but the absorption clearly increases when the observed flux is low. 

\begin{table*}
\centering
\caption{Results of the spectral fits using models $tbabs\times phabs \times \sum^2 apec$ with temperatures fixed to $kT_1=0.75$\,keV and $kT_2=1.9$\,keV. }
\label{fits}
\begin{tabular}{lccccccc}
\hline\hline
ID  & $N_{\rm H}$ & $norm_1$ & $norm_2$ & $\chi_{\nu}^2$ (dof) & $F^{\rm obs}_{\rm X}$ & $F^{\rm unabs}_{\rm X,soft}$ & $F^{\rm unabs}_{\rm X,hard}$\\
    & (10$^{22}$\,cm$^{-2}$) & ($10^{-3}$\,cm$^{-5}$) & ($10^{-3}$\,cm$^{-5}$) &  & \multicolumn{3}{c}{($10^{-12}$\,erg\,cm$^{-2}$\,s$^{-1}$)} \\
\hline
0896/0200450201 & 0.52$\pm$0.05 & 7.83$\pm$0.56 & 3.90$\pm$0.16 & 1.15 (226)& 2.57$\pm$0.04 & 7.23 & 2.26\\
0901/0200450301 & 0.44$\pm$0.03 & 7.41$\pm$0.34 & 4.16$\pm$0.10 & 1.36 (397)& 2.69$\pm$0.03 & 8.02 & 2.33\\
0906/0200450401 & 0.48$\pm$0.03 & 9.55$\pm$0.35 & 4.49$\pm$0.11 & 1.29 (416)& 3.06$\pm$0.03 & 9.23 & 2.63\\
0911/0200450501 & 0.50$\pm$0.03 & 10.3$\pm$0.44 & 2.82$\pm$0.13 & 1.03 (307)& 2.47$\pm$0.04 & 8.91 & 2.00\\
1353/0505110301 & 0.72$\pm$0.06 & 7.09$\pm$0.51 & 2.30$\pm$0.15 & 1.08 (224)& 1.72$\pm$0.04 & 4.39 & 1.48\\
1355/0505110401 & 0.70$\pm$0.06 & 7.47$\pm$0.51 & 2.54$\pm$0.14 & 0.85 (221)& 1.87$\pm$0.03 & 4.81 & 1.61\\
2114/0677980601 & 0.71$\pm$0.05 & 7.11$\pm$0.44 & 1.74$\pm$0.12 & 1.41 (108)& 1.51$\pm$0.04 & 4.31 & 1.25\\
2625/0740300101 & 0.39$\pm$0.04 & 6.07$\pm$0.41 & 3.22$\pm$0.15 & 1.07 (251)& 2.15$\pm$0.04 & 7.07 & 1.84\\
3089/0780040101 & 0.55$\pm$0.06 & 5.87$\pm$0.46 & 2.52$\pm$0.14 & 0.98 (181)& 1.75$\pm$0.04 & 4.96 & 1.52\\
3097/0793183001 & 0.58$\pm$0.08 & 6.32$\pm$0.78 & 4.30$\pm$0.24 & 1.12 (107)& 2.49$\pm$0.07 & 5.70 & 2.76\\
3176/0800150101 & 0.48$\pm$0.04 & 7.98$\pm$0.44 & 3.20$\pm$0.14 & 1.04 (299)& 2.34$\pm$0.04 & 7.59 & 1.98\\
3273/0801910201 & 0.30$\pm$0.08 & 4.55$\pm$0.72 & 4.52$\pm$0.25 & 1.35 (103)& 2.55$\pm$0.07 & 7.31 & 2.30\\
3280/0801910301 & 0.51$\pm$0.09 & 6.78$\pm$0.95 & 3.84$\pm$0.31 & 1.27 (87) & 2.42$\pm$0.08 & 6.46 & 2.15\\
3284/0801910401 & 0.21$\pm$0.14 & 3.30$\pm$1.05 & 4.63$\pm$0.36 & 1.37 (71) & 2.47$\pm$0.10 & 6.98 & 2.27\\
3288/0801910501 & 0.28$\pm$0.07 & 6.81$\pm$0.89 & 4.47$\pm$0.30 & 1.45 (101)& 2.87$\pm$0.08 & 10.1 & 2.49\\
3294/0801910601 & 0.48$\pm$0.08 & 11.7$\pm$1.3  & 3.69$\pm$0.39 & 1.11 (83) & 3.02$\pm$0.11 & 10.6 & 2.51\\
\vspace*{-2mm}\\
00037920001 & 2.14$\pm$0.74 & 16.2$\pm$5.57 & 0.00$\pm$0.93 & 0.50 (2)  & 1.12$\pm$0.41 & 1.68 & 0.98\\
00031904001 & 0.39$\pm$0.28 & 3.97$\pm$2.24 & 3.99$\pm$0.69 & 1.07 (23) & 2.19$\pm$0.20 & 5.49 & 2.01\\
00031904002 & 0.80$\pm$0.32 & 7.10$\pm$2.91 & 2.41$\pm$0.74 & 1.25 (19) & 1.72$\pm$0.16 & 3.91 & 1.52\\
00031904003 & 0.47$\pm$0.26 & 7.47$\pm$2.93 & 3.75$\pm$0.84 & 1.37 (25) & 2.50$\pm$0.22 & 7.42 & 2.18\\
00031904004 & 0.35$\pm$0.32 & 4.10$\pm$2.25 & 3.46$\pm$0.64 & 0.61 (22) & 2.01$\pm$0.17 & 5.73 & 1.80\\
00031904005 & 1.26$\pm$0.39 & 8.40$\pm$3.40 & 2.13$\pm$0.80 & 1.12 (15) & 1.55$\pm$0.19 & 2.44 & 1.42\\
00032767001 & 0.36$\pm$0.64 & 1.40$\pm$4.24 & 3.62$\pm$1.45 & 1.01 (6)  & 1.69$\pm$0.43 & 3.03 & 1.64\\
00032767002 & 0.77$\pm$0.39 & 6.63$\pm$3.47 & 1.98$\pm$0.95 & 0.30 (11) & 1.52$\pm$0.18 & 3.77 & 1.32\\
00033818044 & 0.50$\pm$0.71 & 8.50$\pm$11.9 & 3.99$\pm$3.75 & 1.86 (3)  & 2.71$\pm$0.72 & 8.01 & 2.36\\
00032767003 & 0.94$\pm$0.22 & 8.53$\pm$2.17 & 2.74$\pm$0.54 & 1.11 (37) & 1.93$\pm$0.13 & 3.85 & 1.73\\
00034282018 & 0.94$\pm$0.57 & 9.30$\pm$7.65 & 4.24$\pm$2.19 & 0.81 (8)  & 2.57$\pm$0.49 & 4.51 & 2.38\\
00093146005 & 0.44$\pm$0.64 & 0.85$\pm$3.20 & 4.79$\pm$1.09 & 1.53 (8)  & 2.07$\pm$0.50 & 2.73 & 2.06\\
00093148003 & 1.41$\pm$0.73 & 7.91$\pm$6.44 & 2.57$\pm$1.63 & 1.42 (5)  & 1.60$\pm$0.33 & 2.06 & 1.51\\
00034282086 & 0.88$\pm$0.69 & 8.52$\pm$9.39 & 3.35$\pm$3.20 & 0.54 (4)  & 2.19$\pm$0.68 & 4.32 & 1.98\\
00034282089 & 0.01$\pm$0.87 & 1.40$\pm$8.29 & 5.72$\pm$2.39 & 0.17 (3)  & 2.81$\pm$0.84 & 7.53 & 2.64\\
00010451001 & 1.05$\pm$0.35 & 12.0$\pm$5.12 & 4.77$\pm$1.20 & 1.50 (18) & 2.97$\pm$0.20 & 4.88 & 2.75\\
00010451002 & 1.38$\pm$0.71 & 6.22$\pm$6.00 & 3.26$\pm$1.33 & 0.61 (9)  & 1.69$\pm$0.24 & 1.86 & 1.66\\
00010451003 & 1.42$\pm$0.74 & 8.60$\pm$5.50 & 1.30$\pm$0.95 & 0.95 (8)  & 1.22$\pm$0.19 & 1.93 & 1.09\\
00010451004 & 0.75$\pm$0.42 & 5.27$\pm$3.19 & 2.86$\pm$0.79 & 1.28 (13) & 1.71$\pm$0.20 & 3.48 & 1.57\\
00010451005 & 0.85$\pm$0.57 & 3.17$\pm$3.19 & 3.92$\pm$0.90 & 1.70 (14) & 1.83$\pm$0.23 & 2.39 & 1.80\\
00010451006 & 1.00$\pm$0.32 & 9.67$\pm$3.89 & 4.13$\pm$0.98 & 1.02 (19) & 2.53$\pm$0.23 & 4.28 & 2.35\\
00010451007 & 1.00$\pm$0.50 & 7.45$\pm$4.91 & 4.42$\pm$0.94 & 0.47 (18) & 2.40$\pm$0.22 & 3.58 & 2.29\\
00010451008 & 0.16$\pm$0.54 & 1.08$\pm$2.80 & 4.30$\pm$0.76 & 1.29 (13) & 2.02$\pm$0.25 & 4.24 & 1.94\\
00010451009 & 0.52$\pm$0.27 & 7.98$\pm$3.13 & 3.29$\pm$0.93 & 0.89 (18) & 2.34$\pm$0.23 & 7.04 & 2.02\\
00010451010 & 1.11$\pm$0.45 & 11.4$\pm$5.33 & 2.49$\pm$1.13 & 0.81 (10) & 2.04$\pm$0.26 & 3.89 & 1.81\\
00094061006 & 0.92$\pm$0.99 & 0.00$\pm$3.95 & 3.32$\pm$1.52 & 1.22 (1)  & 1.24$\pm$1.31 & 0.83 & 1.30\\
00088806001 & 0.05$\pm$0.67 & 3.14$\pm$4.91 & 2.90$\pm$1.33 & 1.64 (6)  & 1.80$\pm$0.35 & 8.17 & 1.55\\
00088807001 & 0.20$\pm$0.46 & 4.65$\pm$4.58 & 3.00$\pm$1.43 & 0.39 (6)  & 1.99$\pm$0.34 & 8.06 & 1.69\\
\vspace*{-2mm}\\
2572  &  0.93$\pm$0.14 & 7.52$\pm$0.87 & 1.68$\pm$0.14 & 1.33 (77) & 1.43$\pm$0.04 & 3.24 & 1.24\\
16659 &  0.59$\pm$0.07 & 4.19$\pm$0.28 & 2.77$\pm$0.07 & 1.01 (722)& 1.62$\pm$0.02 & 3.67 & 1.48\\
\vspace*{-2mm}\\
402030010 & 0.44$\pm$0.04 & 6.49$\pm$0.43 & 4.14$\pm$0.12 & 1.04 (1061) & 2.55$\pm$0.04 & 7.24 & 2.27\\
\hline      
\end{tabular}
\\
\tablefoot{Unabsorbed fluxes are corrected for the interstellar column ($2.04\times10^{22}$\,cm$^{-2}$) only. Errors (found using the ``error'' command for the spectral parameters and the ``flux err'' command for the fluxes) correspond to 1$\sigma$; whenever errors were asymmetric, the largest value is provided here. Observed fluxes are expressed in the 0.5--10.0\,keV band, while ISM-corrected fluxes are provided in the soft (0.5--2.0\,keV) and hard (2.0--10.0\,keV) bands. For each dataset, all available data were fit simultaneously: EPIC-pn, MOS1, and MOS2 for \xmm, zeroth and first orders for \ch, 
XIS-0, 1, and 3 for {\it Suzaku}. }
\end{table*}

One \xmm\ point is clearly an outlier, with a stronger flux (Table \ref{fits}) and a harder spectrum (Table \ref{journal}) than other \xmm\ data. This point corresponds to the exposure taken in Rev. 3097 (ObsID 0793183001). As mentioned in Section 2, \duz\ appears far off-axis in this observation and within stray light from Cyg X-3. Despite a careful selection of background, it therefore appears that the stray light contamination cannot be totally eliminated. Therefore, we decided to discard these data from further consideration. We note that the spectra from the other off-axis observations (ObsID=0801910201--601) appear less contaminated, with results usually within the errors (though they are far from perfect).

The \ch\ and \sw\ data nicely agree with \xmm, despite some remaining cross-calibration uncertainties (and a much larger noise for \sw), but the {\it Suzaku} results appear discrepant, especially in flux and $norm_2$ (the difference is less significant for the other parameters and the hardness ratio also seems in agreement). That {\it Suzaku} dataset presented neither an obvious stray light problem nor any reduction trouble. Furthermore, the closest \xmm\ and \sw\ datasets, taken $\sim$200\,d before and after the {\it Suzaku} dataset, respectively, agree well with the overall behaviour of \duz; the origin of the {\it Suzaku} discrepancy therefore remains uncertain.

\section{Optical results}

\subsection{Light-curve analysis}
\citet{got78} found \duz\ to be photometrically variable by up to 0.3\,mag over decades while \citet{lau12} and \citet{sal14} reported changes on shorter timescales. \citet{lau12} found no clear periodicity, only a typical variation timescale of about 30\,d. They also detected a potential trend in $V-I$ colour, possibly linked to some spectral type variations though systematic errors could not be excluded. In contrast, \citet{sal14} detected significant changes with an amplitude in the $I$ filter of 0.18\,mag and with a period of 54.0$\pm$0.1\,d. This period is half the period found in X-rays, suggesting a common origin with two photometric events per X-ray period. 

\begin{figure}
  \begin{center}
\includegraphics[width=8.6cm, bb=1 145 590 290, clip]{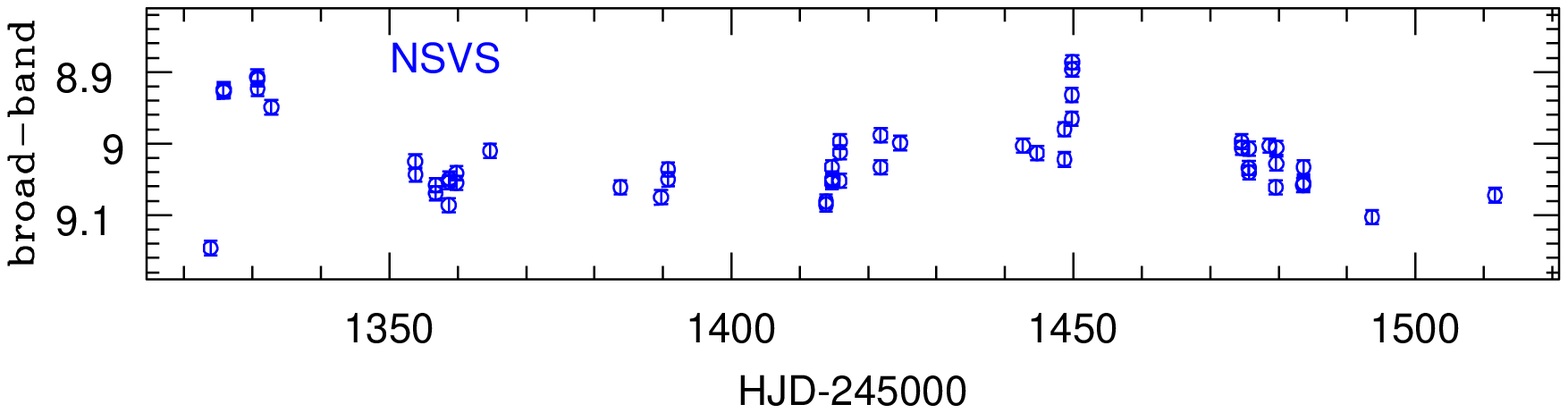}
\includegraphics[width=8.5cm]{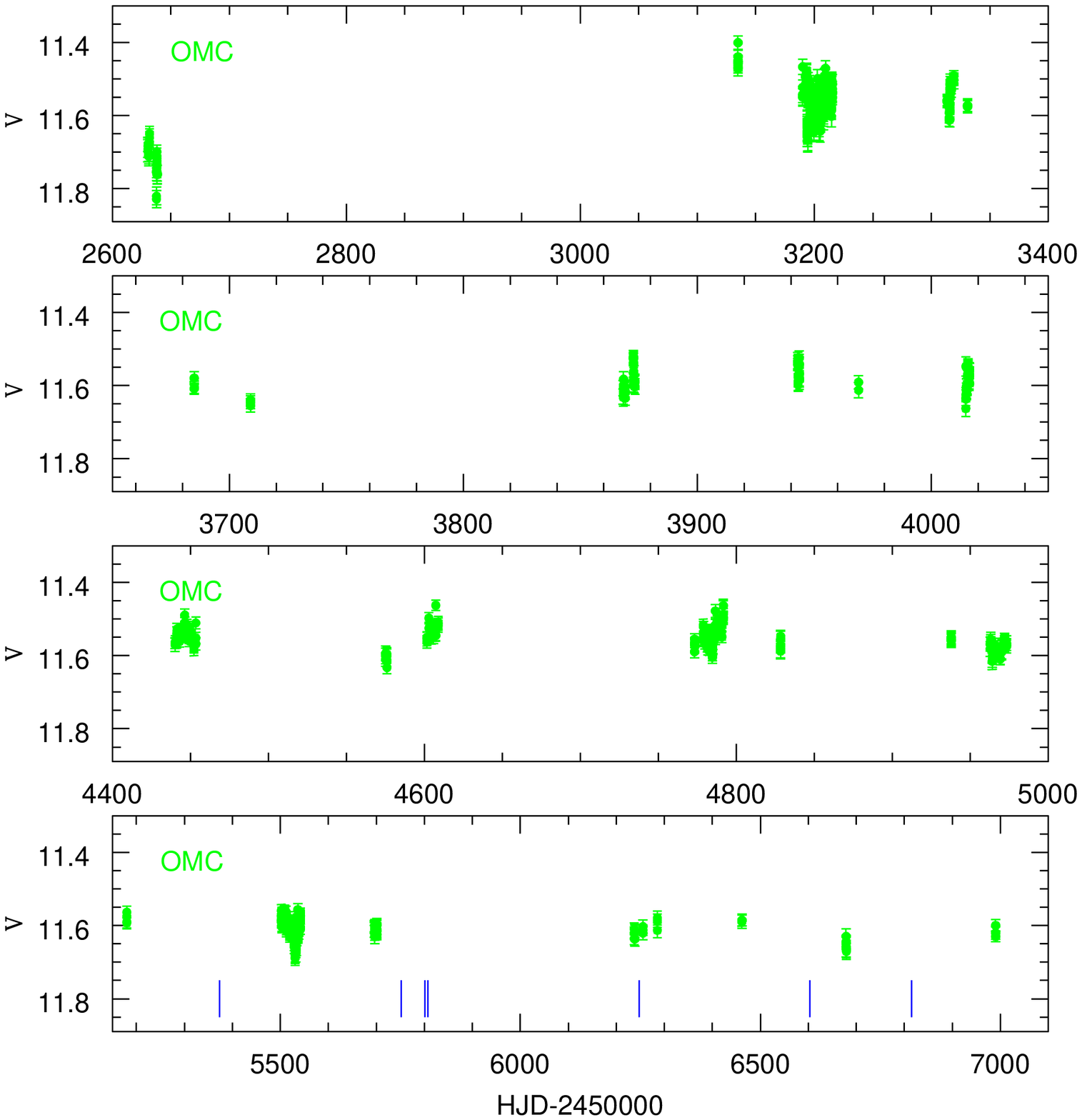}
\includegraphics[width=8.5cm]{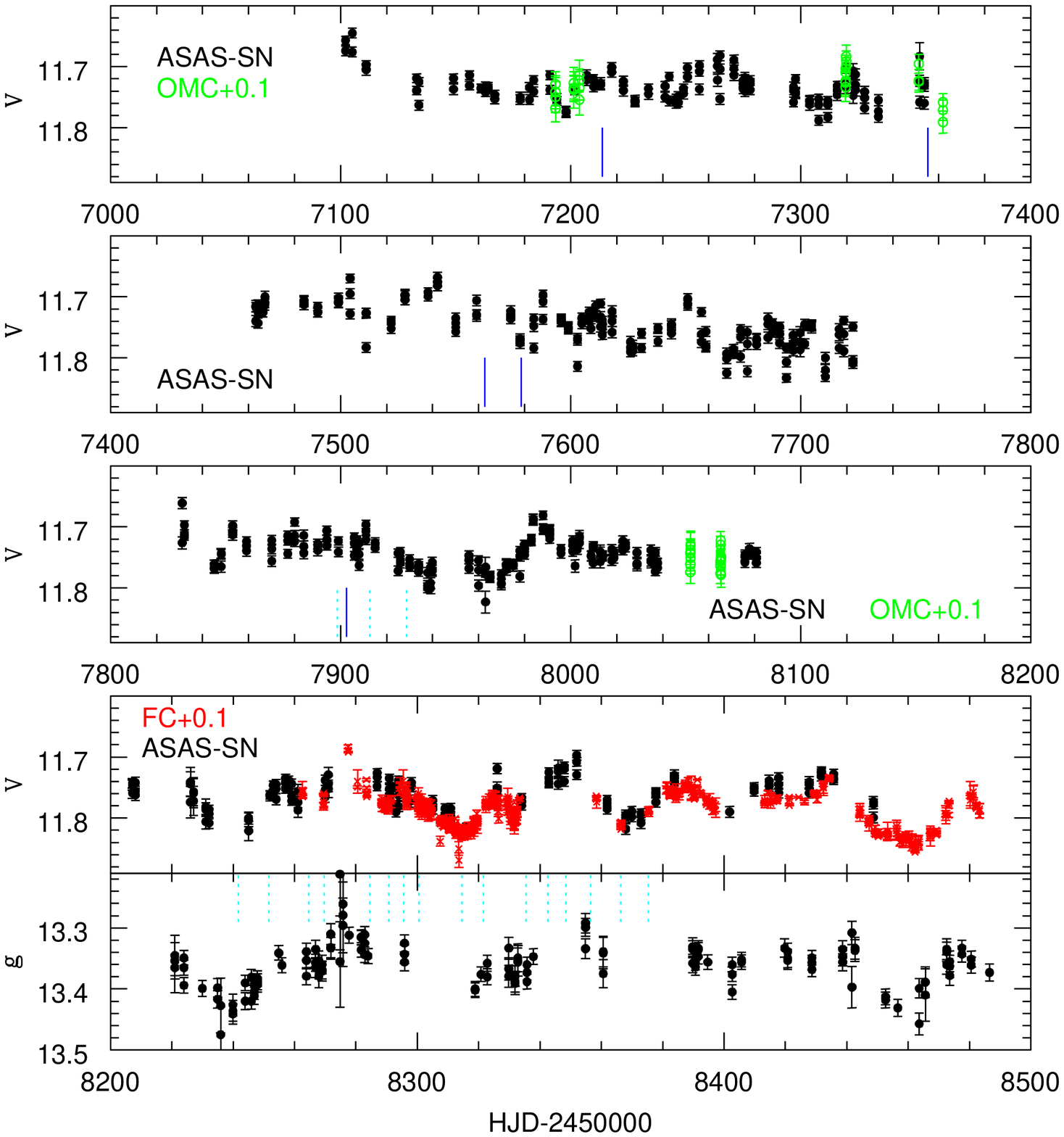}
  \end{center}
\caption{Photometry recorded by one of the present authors (F.C., red): ASAS-SN (black), OMC (green), and NSVS (blue). The times of the Hermes and Carmenes spectroscopic observations are indicated by solid blue and dashed cyan vertical tick symbols, respectively. }
\label{photom}
\end{figure}

\begin{figure*}
  \begin{center}
\includegraphics[width=9cm, bb=20 200 580 720, clip]{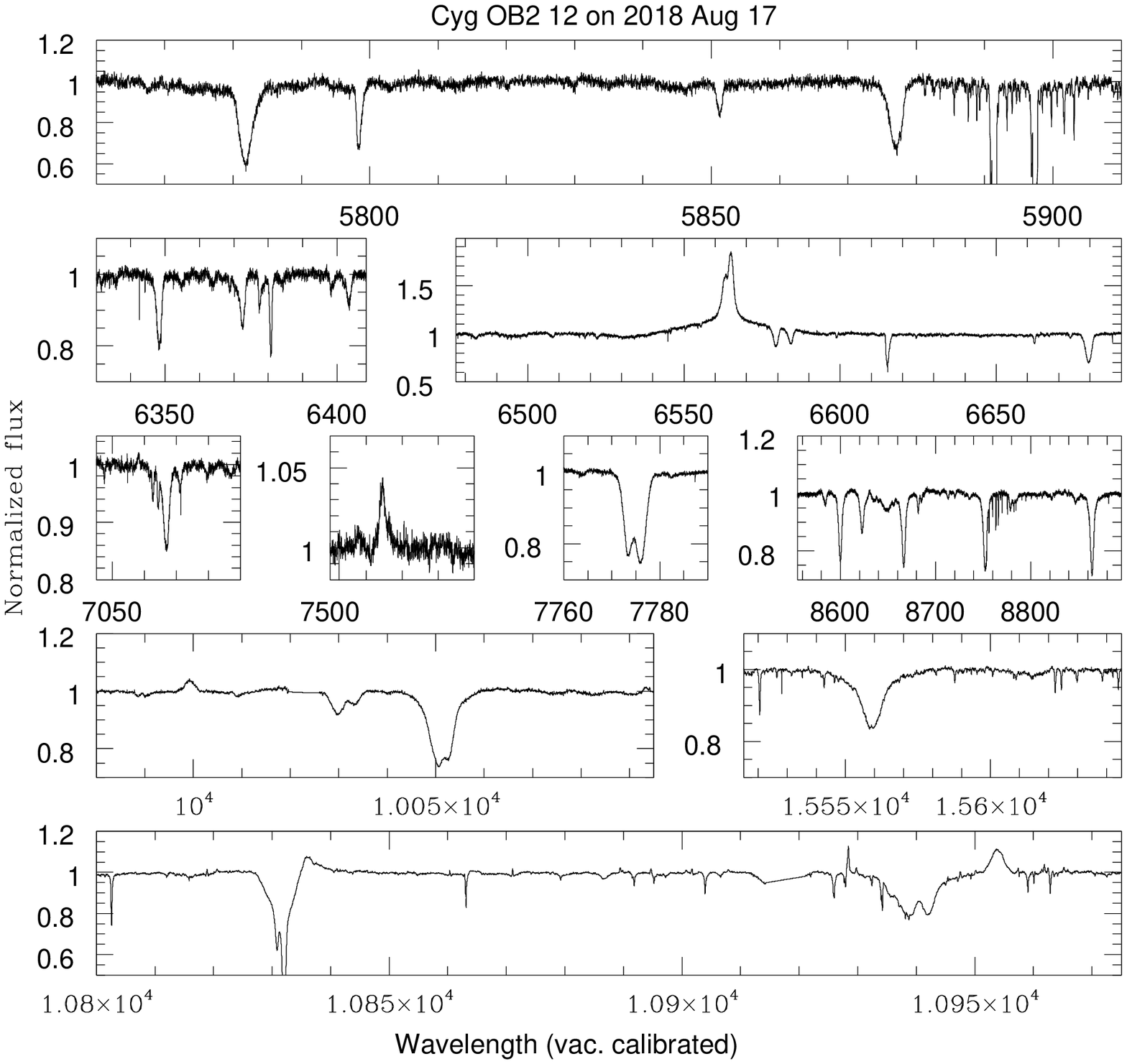}
\includegraphics[width=9cm]{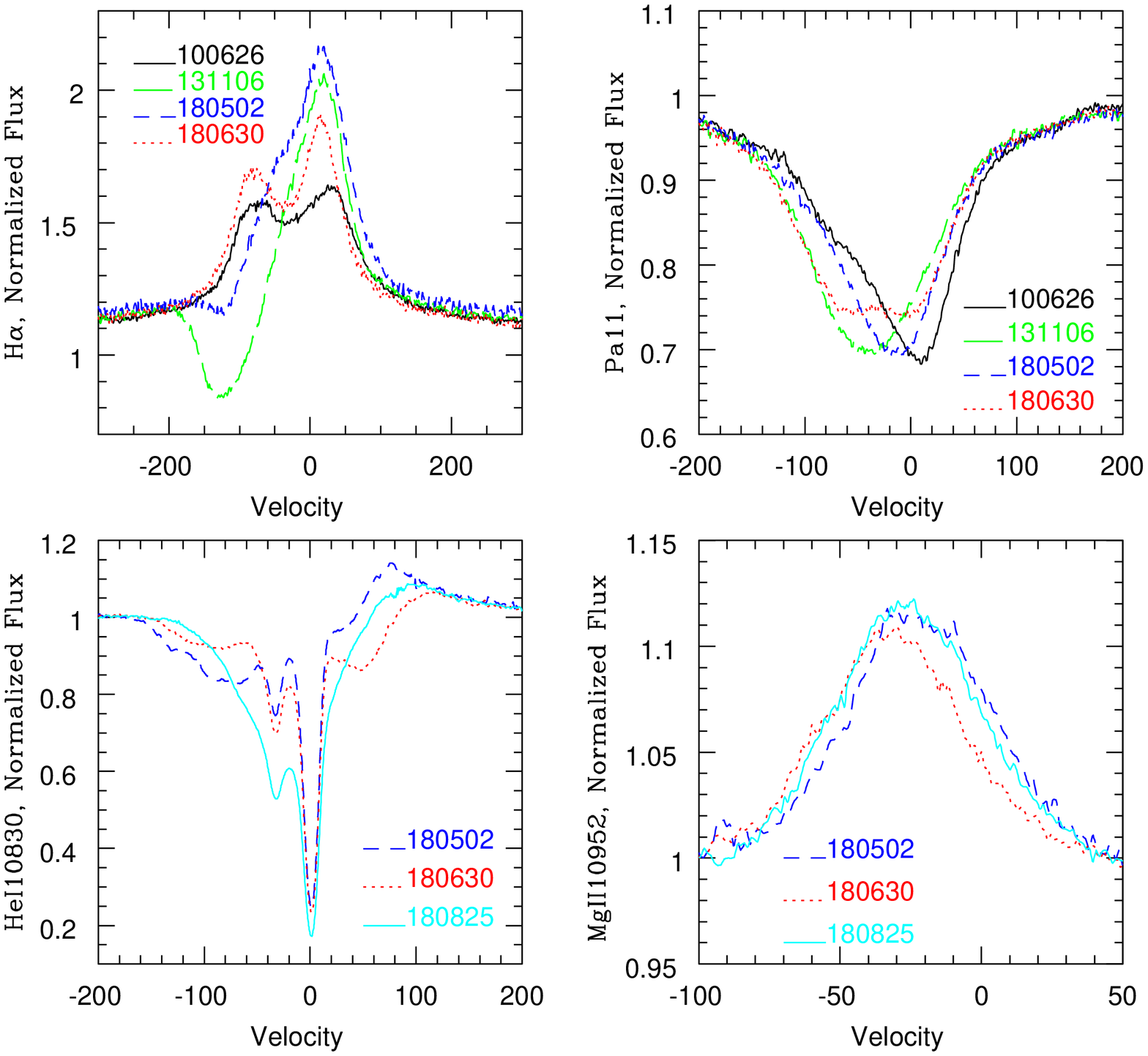}
  \end{center}
\caption{{\it Left:} Excerpts of the yellow-red and near-IR spectrum of \duz, showing its main lines. {\it Right:} Examples of line profile variations. }
\label{spec}
\end{figure*}

The picture is however more complex when one looks at larger datasets. Our recent monitoring of the star was performed at the same time as the spectroscopic campaign (Fig. \ref{photom}). It clearly confirms the brightness variations of \duz, with minima occurring at $HJD\sim2\,458\,315$, 2\,458\,370, and 2\,458\,460.

Public photometric databases provide additional data. The noise affecting Hipparcos photometric measurements and Palomar Transient Finder (PTF) data renders them useless for detecting the small variations of \duz, but significant variations are found in the Northern Survey for Variable Stars (NSVS)\footnote{https://skydot.lanl.gov/nsvs/nsvs.php} and in {\it Integral}-OMC\footnote{http://sdc.cab.inta-csic.es/omc/} observations, though their time coverage is scarce (Fig. \ref{photom}).

ASAS-SN data\footnote{https://asas-sn.osu.edu/} \citep{sha14,koc17} in the $g$ and $V$ filters taken on similar dates to our monitoring agree well with it (as well as with OMC data). The ASAS-SN  data further reveal previous minima around $HJD\sim2\,458\,240$ and possibly 2\,457\,950 (Fig. \ref{photom}). While the separation between the two minima in Summer 2018 is about 54\,d, the previous minima occurred $\sim$75\,d and 365\,d earlier than $HJD\sim2\,458\,315$. In addition, the minimum of Fall 2018 occurred 145\,d later than that date, so the interval between minima does not seem to be constant. Using the X-ray ephemeris, these five minima correspond to phases of 0.86, 0.55, 0.24, 0.75, and 0.59 (considering the dates in chronological order). Only the third one ($HJD\sim2\,458\,315$) appears close to an X-ray minimum. Moreover, the photometric modulation seems to disappear at even earlier dates ($HJD\sim2\,457\,000-800$), with a long monotonic decrease detected between $HJD\sim2\,457\,500$ and $HJD\sim2\,457\,800$. No change in the ASAS-SN camera or observing strategy was performed however, meaning that this is not an instrumental effect. \citet{sal14} reported their detection of a periodic signal for observations taken in $HJD\sim2\,456\,100-600$. It therefore seems that the periodic signal may be transient. 

Using the same methods as before, we performed a period search on the two largest datasets (ASAS-SN and OMC in $V$; see left panels of Fig. \ref{addps}). The periodogram based on ASAS-SN data only shows a peak at very low frequencies, at a position close to a peak in the spectral window: no significant periodicity is therefore detected (as could be expected from the discussion above). On the other hand, the OMC periodogram possesses a clear peak at about 87\,d, but the scarce time sampling does not fully cover a single 87\,d period (i.e. there are no set of 5--10 points taken within 87\,d): it is therefore difficult to fully confirm it, and this signal could actually be related to the 54--108\,d timescales. We also performed a period search on the most recent data only (last run of ASAS-SN data and our own photometric campaign; see right panels of Fig. \ref{addps}) since this is when the oscillations are most apparent. While they cover similar time intervals (see Fig. \ref{photom}), though with different sampling, the two datasets do not yield the exact same periodograms, with the highest peak at 73.6$\pm$2.2\,d for ASAS-SN and at 48.4$\pm$1.1\,d or 130$\pm$8\,d for our own data. This confirms the probable absence of significant strict periodicity in the photometry, though significant variations on timescales of tens of days are clearly present.

\subsection{Spectroscopic features}

Apart from a large number of strong diffuse interstellar bands \citep[DIBs, e.g.\ 5780, 5797, 6613\,\AA,][]{Herbig}, the optical and near-IR spectrum of \duz\ contains prominent hydrogen and He\,{\sc i} lines, but also a few metallic lines (Fig. \ref{spec}). In particular, Paschen and Brackett hydrogen lines, He\,{\sc i}\,$\lambda\lambda$\,5876,\,6678,\,7065\AA, Si\,{\sc ii}(+Mg\,{\sc ii})\,$\lambda$\,6347\AA, Si\,{\sc ii}\,$\lambda$\,6371\AA, Ne\,{\sc i}\,$\lambda$\,6402\,\AA, N\,{\sc ii}\,$\lambda$\,6482\,\AA, C\,{\sc ii}\,$\lambda\lambda$\,6578,\,6583\AA, O\,{\sc i}\,$\lambda\lambda$\,7772,\,7774,\,8446\AA, and N\,{\sc i}\,$\lambda\lambda$\,8680--8718\AA\  appear in absorption, while He\,{\sc i}\,$\lambda$\,10830\AA\ displays an incipient P-Cygni profile, H$\alpha$ displays a strong emission, and Fe\,{\sc ii}\,$\lambda$\,7513,\,9998\AA\ and Mg\,{\sc ii}\,$\lambda$\,10952\AA\ show weaker emissions.

\begin{figure*}
  \begin{center}
\includegraphics[width=4.5cm]{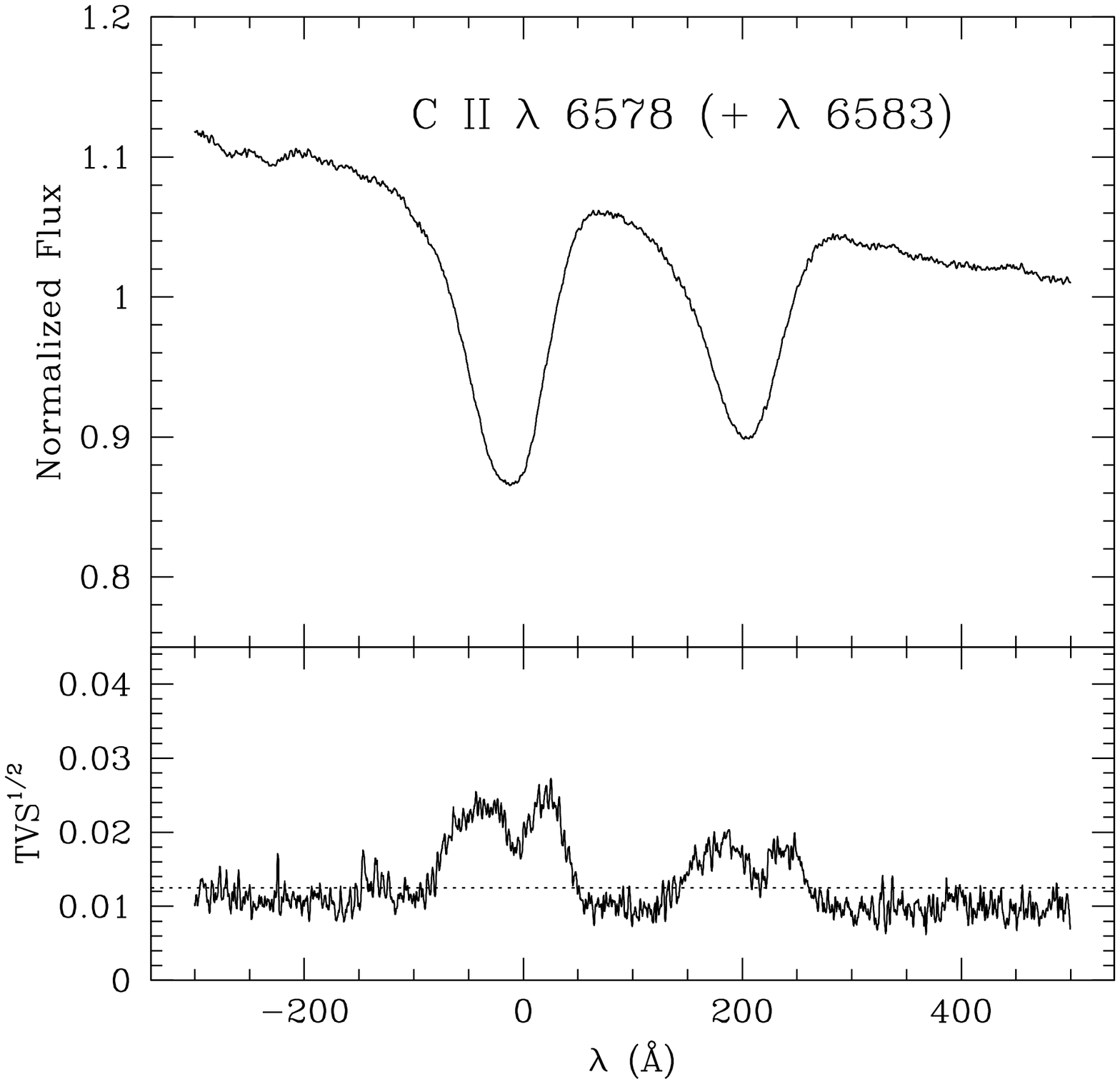}
\includegraphics[width=4.5cm]{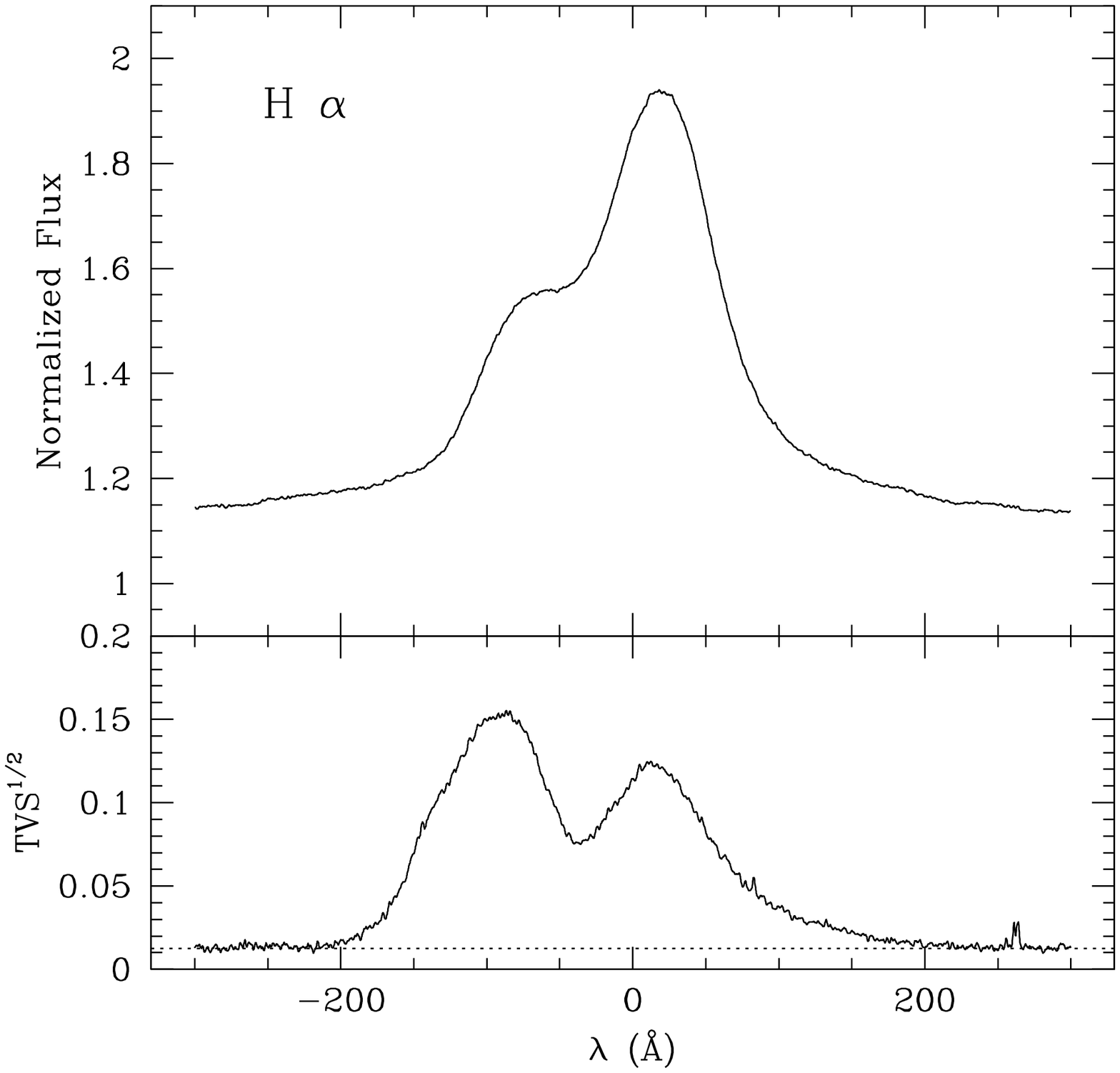}
\includegraphics[width=4.5cm]{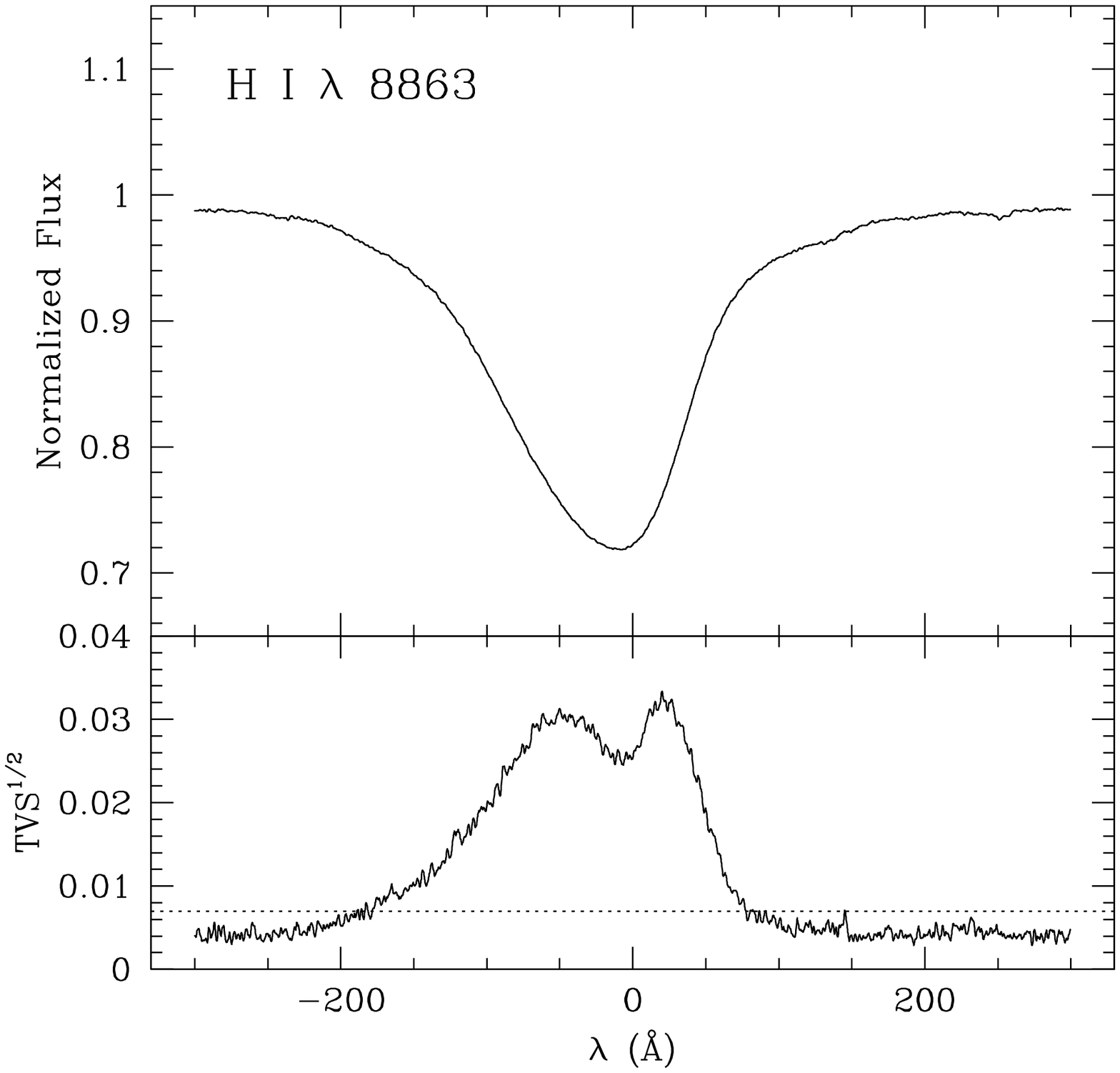}
\includegraphics[width=4.5cm]{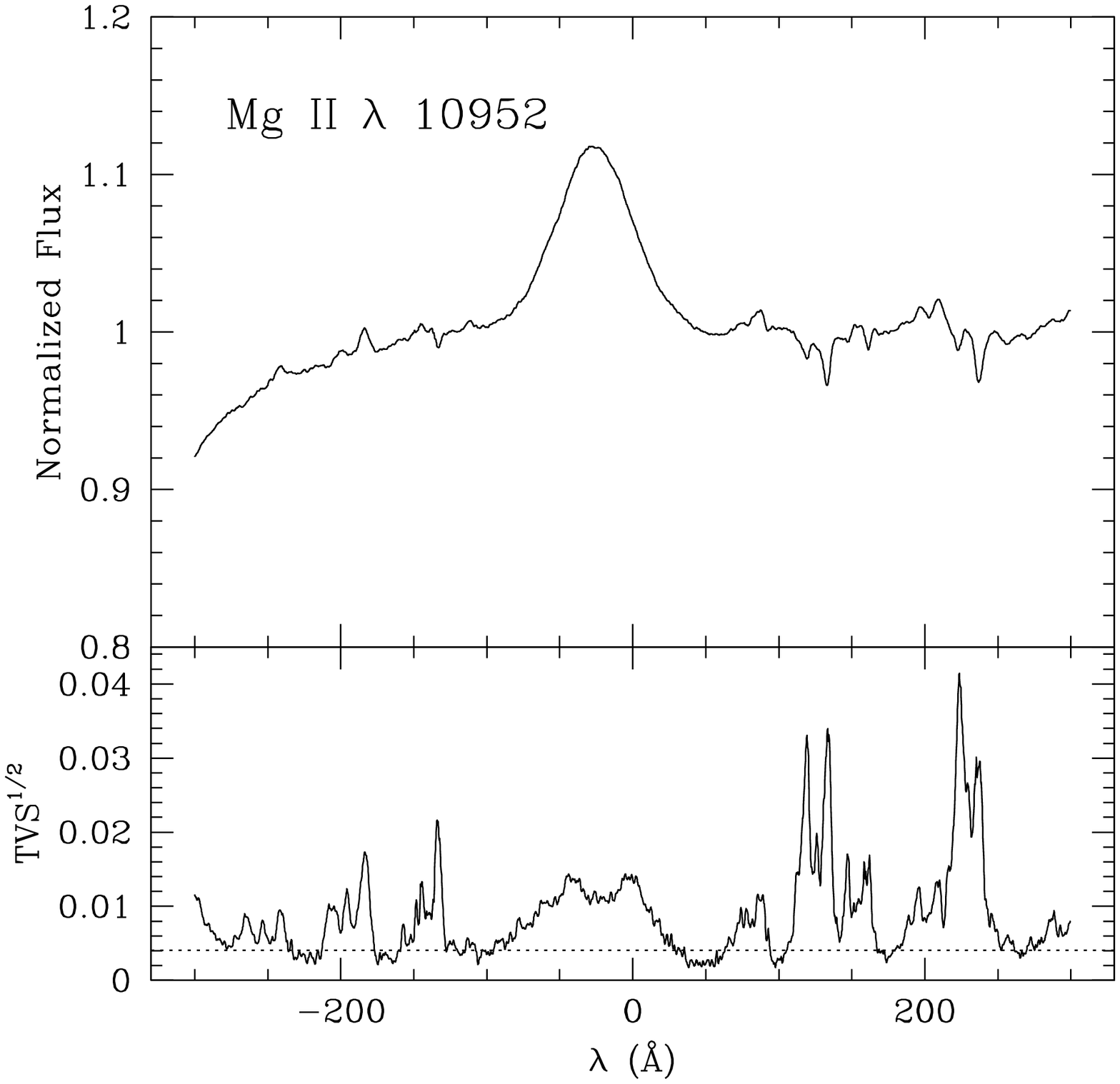}
\end{center}  
  \caption{Mean spectrum and temporal variance spectrum for various lines in the spectrum of \duz. The dotted lines in the TVS$^{1/2}$ panels correspond to the 1\% significance level of variability evaluated accounting for the S/N of the data. On the sides of the Mg\,{\sc ii} line, significant narrow variabilities are due to the imperfect correction of the telluric lines.
\label{tvs}}
\end{figure*}

Criteria for the spectral classification of OB supergiants based on red spectra in the region around H$\alpha$ and in the {\it I}-band were discussed by \citet{Neg10}. In the spectrum of \duz, the rather narrow Paschen lines as well as the relative strength of the O\,{\sc i} $\lambda$\,8446 line compared to the neighbouring H\,{\sc i}\,$\lambda$\,8438 clearly indicate a mid-B supergiant star. On the other hand, the C\,{\sc ii} lines are less constraining since they are strong for giants and especially so for supergiants of spectral types between B1 and B5 \citep{Walborn,Neg10}. Although clearly present, the N\,{\sc ii}\,$\lambda$\,6482\AA\ line is weak, suggesting a spectral type later than B2, which is confirmed by the presence of strong Si\,{\sc ii} lines \citep{Walborn}. The presence of weak but definite N\,{\sc i} lines in the region between 8680 and 8718\,\AA\ indicates a spectral type later than B3 but earlier than B6 \citep{Neg10}. Comparing the equivalent widths ($EW$s) of Si\,{\sc ii} $\lambda$\,6347 and Ne\,{\sc i} $\lambda$\,6402 ($EW = 0.34\pm 0.02$\,\AA\ and $0.16 \pm 0.02$\,\AA, respectively - see Table \ref{averm}) with the results of \citet[Figs. 21 and 24]{Len93} further yields a spectral type around B5 with an uncertainty or a range of variation (see below) of about one subclass. The emission component of the He\,{\sc i}\,$\lambda$\,10830\AA\ line is weaker than what is observed in very luminous (luminosity class Ia$^+$) early B hypergiants \citep{Groh}. The weak Fe\,{\sc ii}\,$\lambda$\,9998\AA\ emission line was seen in the spectra of HD\,183143 (B7\,Ia) and $\mu$\,Sgr (B8\,Iap) presented by \citet{CH}, but is absent in the spectrum of HD\,225094 (B3\,Ia) shown by the same authors. The above considerations suggest a spectral type B5\,Ia for \duz. This result is in reasonable agreement with the B3-4\,Ia$^+$ classification determined by \citet{cla12}, except for the luminosity class. Whereas \citet{cla12} considered \duz\ to be a blue hypergiant, we favour a normal supergiant, which is also in line with the revised distance (see following section). 

We compared the spectra taken close to minimum brightness (2018 May, July, and Sep) with those taken near maximum brightness (2018 June and mid-August). Several lines which are sensitive to spectral type \citep{Len93,Neg10}, and thus temperature, display changes of their strength that seem related to the brightness variations. For instance, the H\,{\sc i} absorptions clearly appear stronger when the star is fainter and its $V - I$ index is higher (Fig. \ref{sptype}). A similar behaviour, though of smaller amplitude, is seen in Ne\,{\sc i}\,$\lambda$\,6402\AA, Si\,{\sc ii}\,$\lambda\lambda$\,6347,\,6371\AA, and N\,{\sc i}\,$\lambda$\,8680\AA. Correlated changes are less obvious for the C\,{\sc ii}\,$\lambda\lambda$\,6578,\,6583\AA,\ and N\,{\sc ii}\,$\lambda$\,6482\AA\ lines, but they are less sensitive to temperature effects. The observed variations suggest changes of the spectral type between about B4 and B6. Such variations of the temperature are consistent with the $V - I$ colour variations of about 0.05\,mag that we found to be correlated with the brightness variations of the star.

Our dataset is composed of 31 high-resolution spectra (of which 19 also cover the near-IR wavelengths) obtained between 2010 and 2018, with a denser monitoring over the last year (one spectrum every $\sim$10\,d). Comparing these spectra, it is clear that line profiles change from one observation to the next (Fig. \ref{spec}). At some point, most lines seem to skew and even become double, a telltale sign of binarity. However, in the process, the depths of these lines do not change significantly, contrary to what is seen in spectroscopic binaries. Furthermore, other lines observed simultaneously display no sign of splitting, such as the Mg\,{\sc ii} and Fe\,{\sc ii} emissions which simply seem to shift. Finally, H$\alpha$ and He\,{\sc i}\,$\lambda$\,10830\AA\ show a more complex behaviour. The H$\alpha$ profile is mostly composed of a broad emission with one or two peaks. In the latter case, both the relative intensity of the two peaks and their separation may vary. In addition, a shallow blueshifted absorption is sometimes (but not always) present. These variations are much more extreme than those reported by \citet{che13} and \citet{klo04}. Finally, the He\,{\sc i}\,$\lambda$\,10830\AA\ line profile is composed of three parts: a constant interstellar component (two deep and narrow absorption lines at velocities $\sim$0 and --30\kms), a broad emission component on the red wing, and a complex absorption on both wings. When overplotting all data, the broad emission appears constant, whereas the absorption features vary in number, depth, width, and position.

To further characterise the variability, we computed the temporal variance spectrum \citep[TVS,][]{FGB96} for the main lines mentioned above (see a few examples in Fig. \ref{tvs}). To this aim, we combined the Hermes and Carmenes spectral sets, which can be done in view of their similar resolving powers and of the fact that the line widths are largely dominated by the star, not the instrument. The TVS confirms the presence of significant variations over a large range of velocities, larger than the $-v\,\sin{i}$ to $+v\,\sin{i}$ range \citep[$v\,\sin{i}=38$\,\kms, see][]{cla12}. For all examined lines, the TVS displays a double-peaked structure. Such a profile is expected if the line centroid shifts back and forth in radial velocity. This can be the result of a genuine binary, pulsations, or co-rotating large-scale structures in the stellar wind. Another remarkable feature can be found when examining the symmetry of the TVS: the blue wing becomes progressively stronger when going from lines formed deep in the photosphere (such as the S\,{\sc ii} or C\,{\sc ii} lines which have roughly symmetric TVS profiles) to lines arising farther out (such as the hydrogen Paschen lines which have blue wings extending out to $-150$\,km\,s$^{-1}$). This probably reflects increasing contributions from wind material to the absorption. The TVS of the H$\alpha$ emission line reveals the largest and broadest variability (up to 15\% of the continuum, between --200\,\kms\ and 150\,\kms), with He\,{\sc i}\,$\lambda$\,10830\AA\ being second (variations up to 9\% in --150\,\kms\ and 100\,\kms), but even the quieter Mg\,{\sc ii} emission exhibits changes by more than 1\%. 

To characterise the lines, fitting profiles of a given shape (e.g. Gaussian) is a usual method. However, because of the variability of the profiles, it is quite difficult to obtain a good fit over the full profile - only meaningful radial velocities can be derived by fitting the bottom (resp. top) of the absorption (resp. emission) lines. Therefore, we decided to focus on line moments. Subtracting 1 from the normalized spectra, we calculated the zeroth-order moment $M_0=\sum (F_i-1)$ which, multiplied by the wavelength step, yields the $EW$, the first-order moment $M_1=\sum (F_i-1)\times v_i /\sum (F_i-1),$ which provides the centroid, and the second-order moment $M_2=\sum (F_i-1)\times (v_i-M_1)^2 /\sum (F_i-1),$ whose square root gives the line width. Because of the noise, we avoided considering further orders. For each line, we limit the calculation to the wavelength interval where the line varies, using most of the profile while avoiding nearby lines. Table \ref{averm} displays the averaged values of these moments ($<X>=\sum X_i/N$), along with their dispersions ($\sqrt {\sum (X_i-<X>)^2/(N-1)}$). All data were used in these averages, though a few noisy spectra (see Table \ref{jouropt}) clearly add to the dispersion, especially for the bluer wavelengths. A very good agreement is found when comparing the two datasets, Carmenes and Hermes, demonstrating the reliability of the reduction. The diffuse interstellar bands (DIBs) enable us to check the (small) ``natural'' noise around the moment values, due to the number of collected photons and the quality of the reduction (wavelength calibration, normalization, etc.). This provides a threshold against which variability can be detected. Equivalent widths and widths generally are rather stable except for the two lines strongly affected by wind, H$\alpha$ and He\,{\sc i}\,$\lambda$\,10830\AA. It is interesting to note that the dispersions in $EW$ and width reach their smallest values not only for DIBs but also for Mg\,{\sc ii} and Fe\,{\sc ii} lines: this confirms that these stellar lines vary very little in shape. The centroids are however found to vary for all stellar lines, with the largest dispersions found for the wind lines and the smallest (again) for DIBs, and then for Mg\,{\sc ii} and Fe\,{\sc ii} emission lines. 

\begin{table*}
\centering
\caption{Average moments for different lines. }
\label{averm}
\begin{tabular}{lcccccc}
  \hline\hline
Line & $\lambda_0$ & limit$_{blue}$ & limit$_{red}$ & $EW$ & centroid & width \\
     & \AA\  & \kms\ & \kms\ & \AA\ & \kms\ & \kms\ \\
\hline
DIB         & 5780.450 (a) & --150 & 150 & 1.09$\pm$0.04 &  --3.2$\pm$1.9 & 60.7$\pm$1.8  \\
DIB         & 5782.053 (v) & --150 & 150 & 1.03$\pm$0.08 &  --3.3$\pm$1.3 & 57.4$\pm$4.1  \\
\vspace*{-2mm}\\
He\,{\sc i} & 5875.621 (a) & --150 & 150 & 0.62$\pm$0.06 & --20.2$\pm$8.1 & 34.4$\pm$10.0 \\
He\,{\sc i} & 5877.249 (v) & --150 & 150 & 0.70$\pm$0.05 & --15.4$\pm$4.6 & 45.6$\pm$5.6  \\
\vspace*{-2mm}\\
Si\,{\sc ii}& 6347.11 (a)  & --100 & 50  & 0.34$\pm$0.02 & --26.6$\pm$4.8 & 31.8$\pm$1.8 \\
Si\,{\sc ii}& 6348.86 (v)  & --100 & 50  & 0.34$\pm$0.03 & --26.6$\pm$3.6 & 32.0$\pm$1.4 \\
\vspace*{-2mm}\\
Si\,{\sc ii}& 6371.37 (a)  & --100 & 50  & 0.24$\pm$0.02 & --29.4$\pm$4.7 & 32.6$\pm$2.2 \\
Si\,{\sc ii}& 6373.13 (v)  & --100 & 50  & 0.24$\pm$0.02 & --29.5$\pm$3.7 & 32.6$\pm$2.0 \\
\vspace*{-2mm}\\
Ne\,{\sc i}  & 6402.248 (a) & --150 & 50 & 0.16$\pm$0.03 & --32.1$\pm$4.5 & 40.7$\pm$3.0  \\
Ne\,{\sc i}  & 6404.018 (v) & --150 & 50 & 0.16$\pm$0.02 & --33.0$\pm$4.2 & 39.9$\pm$2.5  \\
\vspace*{-2mm}\\
N\,{\sc ii}  & 6482.05 (a) & --100 & 50 & 0.088$\pm$0.009 & --22.1$\pm$7.0 & 30.0$\pm$6.0  \\
N\,{\sc ii}  & 6483.84 (v) & --100 & 50 & 0.080$\pm$0.016 & --22.9$\pm$6.1 & 32.8$\pm$2.4  \\
\vspace*{-2mm}\\
H\,{\sc i}  & 6562.85 (a)  & --300 & 300 &--4.54$\pm$0.63& --2.1$\pm$10.6 &124.9$\pm$4.7  \\
H\,{\sc i}  & 6564.66 (v)  & --300 & 300 &--4.51$\pm$0.35&  --4.2$\pm$8.3 &126.4$\pm$2.8  \\
\vspace*{-2mm}\\
C\,{\sc ii} & 6578.05 (a)  & --110 & 110 & 0.37$\pm$0.02 & --17.7$\pm$4.3 & 35.7$\pm$1.8  \\
C\,{\sc ii} & 6579.87 (v)  & --110 & 110 & 0.37$\pm$0.02 & --19.2$\pm$3.8 & 34.9$\pm$2.0  \\
\vspace*{-2mm}\\
C\,{\sc ii} & 6582.88 (a)  & --110 & 110 & 0.25$\pm$0.02 & --25.2$\pm$6.0 & 29.3$\pm$4.6  \\
C\,{\sc ii} & 6584.70 (v)  & --110 & 110 & 0.25$\pm$0.02 & --25.8$\pm$4.6 & 30.1$\pm$4.2  \\
\vspace*{-2mm}\\
DIB         & 6613.62 (a)  & --70  & 70  &0.372$\pm$0.007&  --5.3$\pm$0.4 & 20.9$\pm$0.9  \\
DIB         & 6615.447(v)  & --70  & 70  &0.386$\pm$0.014&  --5.0$\pm$0.5 & 21.6$\pm$1.3  \\
\vspace*{-2mm}\\
He\,{\sc i} & 6678.151 (a) & --130 & 130 & 0.69$\pm$0.05 & --16.1$\pm$6.1 & 41.0$\pm$1.8  \\
He\,{\sc i} &6679.9949 (v) & --130 & 130 & 0.70$\pm$0.04 & --17.3$\pm$4.4 & 41.7$\pm$1.9  \\
\vspace*{-2mm}\\
He\,{\sc i} & 7065.19 (a)  & --90  & 90  & 0.34$\pm$0.04 &  --6.0$\pm$7.3 & 36.6$\pm$2.1  \\
He\,{\sc i} & 7067.138 (v) & --90  & 90  & 0.34$\pm$0.04 &  --5.0$\pm$4.9 & 37.5$\pm$1.7  \\
\vspace*{-2mm}\\
Fe\,{\sc ii}& 7513.161 (a) &--75   & 25  &--0.061$\pm$0.005 & --23.1$\pm$2.0 & 23.9$\pm$0.8 \\
Fe\,{\sc ii}& 7515.230 (v) &--75   & 25  &--0.062$\pm$0.007 & --23.8$\pm$2.6 & 23.9$\pm$0.8 \\
\vspace*{-2mm}\\
H\,{\sc i}  & 8598.392 (a) & --150 & 150 & 1.05$\pm$0.04 & --21.3$\pm$5.1 & 60.6$\pm$2.0  \\
H\,{\sc i}  & 8600.754 (v) & --150 & 150 & 1.04$\pm$0.04 & --21.5$\pm$3.6 & 59.8$\pm$1.6  \\
\vspace*{-2mm}\\
N\,{\sc i}  & 8680.282 (a) & --100 & 40 & 0.14$\pm$0.02 & --26.8$\pm$4.6 & 29.1$\pm$2.7  \\
N\,{\sc i}  & 8682.666 (v) & --100 & 40 & 0.14$\pm$0.02 & --27.2$\pm$4.2 & 30.0$\pm$1.9    \\
\vspace*{-2mm}\\
N\,{\sc i}  & 8711.703 (a) & --100 & 40 & 0.031$\pm$0.011 & --36.1$\pm$9.1 & 19.9$\pm$10.1  \\
N\,{\sc i}  & 8714.096 (v) & --100 & 40 & 0.026$\pm$0.014 & --34.6$\pm$10.1 & 27.9$\pm$18.5   \\
\vspace*{-2mm}\\
H\,{\sc i}  & 8862.783 (a) & --150 & 150 & 1.32$\pm$0.06 & --19.8$\pm$4.8 & 62.7$\pm$1.7  \\
H\,{\sc i}  & 8865.217 (v) & --150 & 150 & 1.33$\pm$0.05 & --19.3$\pm$4.0 & 62.9$\pm$0.9  \\
\vspace*{-2mm}\\
Fe\,{\sc ii}& 10000.32 (v) & --75  & 25  &--0.083$\pm$0.011& --25.5$\pm$2.6 & 23.9$\pm$0.6 \\
\vspace*{-2mm}\\
H\,{\sc i}  &10052.128 (v) & --150 & 150 & 1.57$\pm$0.08 & --20.6$\pm$4.8 & 63.3$\pm$1.4  \\
\vspace*{-2mm}\\
He\,{\sc i} &10832.0574(v) & --170 & 170 & 1.10$\pm$0.20 & --48.6$\pm$20.4& 52.6$\pm$19.5 \\
\vspace*{-2mm}\\
Mg\,{\sc ii}& 10954.77 (v) & --75  & 25  &--0.26$\pm$0.02& --25.5$\pm$3.0 & 22.4$\pm$0.6  \\
\vspace*{-2mm}\\
H\,{\sc i}  &15560.699 (v) & --150 & 150 & 1.52$\pm$0.08 & --16.7$\pm$4.1 & 67.0$\pm$1.7  \\
\hline
\end{tabular}
\\
\tablefoot{Values are provided for the two sets of data separately, which is indicated in the rest wavelength column: Carmenes spectra were calibrated in vacuum  (v) and Hermes data in the air (a) - usually wavelengths were taken from http://www.pa.uky.edu/$\sim$peter/atomic/. $EW$s are negative for emission lines and positive for absorptions. As the C\,{\sc ii} lines are in the red emission wing of H$\alpha$, moments were calculated considering a local continuum accounting for the emission wing.}
\end{table*}

\begin{figure}
\includegraphics[width=8.5cm]{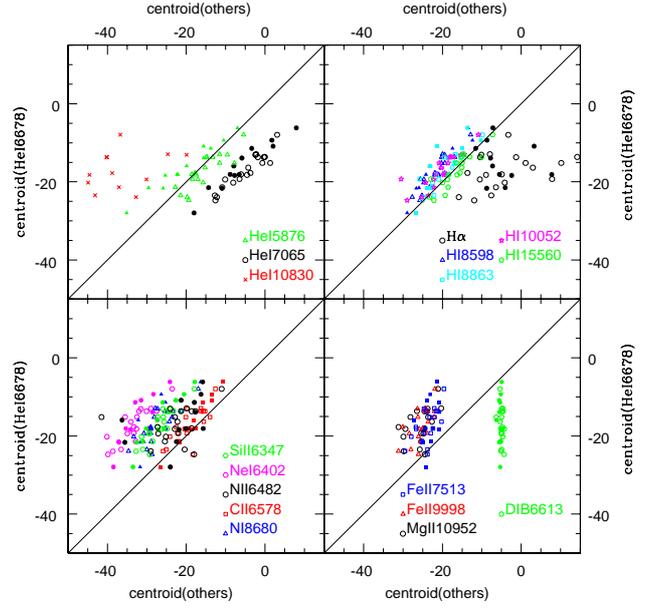}
\caption{Comparisons between the centroid of He\,{\sc i}\,$\lambda$\,6678\AA\ and the centroids of other lines detected in the spectrum of \duz\ (He\,{\sc i} lines on top left, H\,{\sc i} lines on top right, and metallic absorption and emission lines on bottom left and right, respectively). Filled symbols correspond to Hermes data, open symbols and crosses to Carmenes data; the diagonal yields the one-to-one correlation. For comparison, the case of a DIB is added to the last panel.}
\label{correlm}
\end{figure}

To clarify the behaviour of the lines, we compared the individual centroids of different lines in Fig. \ref{correlm}. Except for the wind lines, strong correlations between centroids are found. But the devil lies in the details. Indeed, the amplitudes of variations fall into two groups: the centroids of He\,{\sc i}\,$\lambda\lambda$\,5876,\,6678,\,7065\AA, Si\,{\sc ii}\,$\lambda\lambda$\,6347,\,6371\AA, C\,{\sc ii}\,$\lambda\lambda$\,6578,\,6583\AA, and H\,{\sc i}\,$\lambda\lambda$\,8598,\,8863,\,10052,\,15560\AA\ (all absorption lines) vary by about 25\kms\ peak-to-peak, whereas those of Fe\,{\sc ii}\,$\lambda\lambda$\,7513,\,9998\AA,\ and Mg\,{\sc ii}\,$\lambda$\,10952\AA\ emissions only change between --20 and --30\kms. Indeed, a unity slope can be used for all cases except when comparing centroids from the first group with those of the second (see bottom right panel of Fig. \ref{correlm}). However, it is also clear that even the variations of Fe\,{\sc ii} and Mg\,{\sc ii} are much larger than that observed for DIBs. 

\begin{figure}
\includegraphics[width=8.5cm]{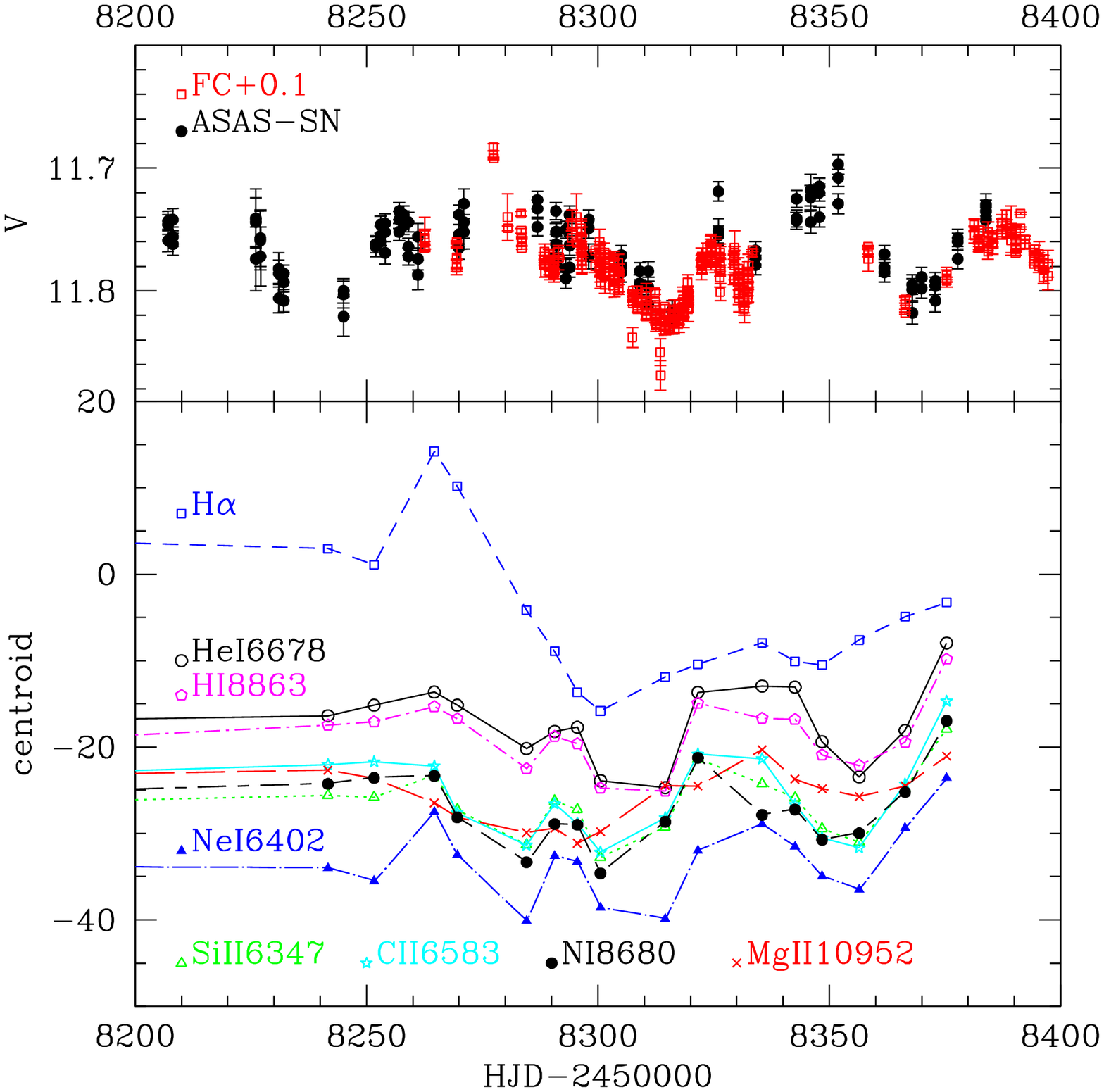}
\caption{Evolution of the photometry of \duz\ and of the centroids of some prominent lines in its spectrum during the Carmenes campaign in Summer 2018. }
\label{rvphot}
\end{figure}

\citet{Rivinius} reported the results of an intense monitoring of three early-B hypergiants. All three stars displayed significant velocity variations of their photospheric lines. For instance, in the case of $\zeta^1$\,Sco, \citet{Rivinius} observed a pulsation-like pattern with a peak-to-peak amplitude of 20\,\kms on timescales near 10--15\,d that gave way after some time to a more random variation. Lines originating from outer layers of the photosphere were found to display a delayed variation pattern compared to lines formed in deeper layers. In addition, we may note that photometric variations with an amplitude near 0.1\,mag were observed for $\zeta^1$\,Sco on the same timescales \citep{Burki}. A similar situation is observed in the case of \duz\ (though with a longer variation timescale, 50--100\,d): the velocities of He\,{\sc i}\,$\lambda$\,6678\AA\ and H\,{\sc i}\,$\lambda$\,8863\AA\ lines display the same qualitative behaviour as those of C\,{\sc ii}\,$\lambda\lambda$\,6578,\,6583\AA\ lines, but with a slight delay of the order of 3--5\,d (see Fig. \ref{rvphot}). 

\begin{figure}
\includegraphics[width=8.5cm]{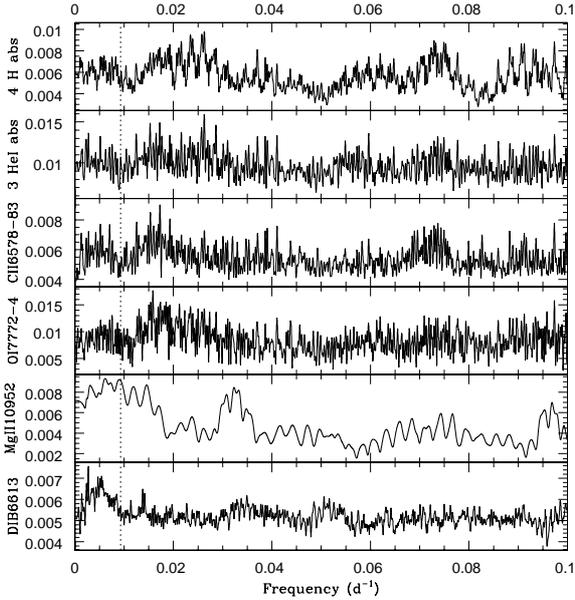}
\caption{Periodograms calculated on the line profiles (see text) for H\,{\sc i}\,$\lambda\lambda$\,8598,\,8663,\,10052,\,15560\AA, He\,{\sc i}\,$\lambda\lambda$\,5876,\,6678,\,7065\AA, C\,{\sc ii}\,$\lambda\lambda$\,6578,\,6583\AA, O\,{\sc i}\,$\lambda\lambda$\,7772,\,7774\AA, Mg\,{\sc ii}\,$\lambda$\,10952\AA, and the DIB at 6613\AA\ for comparison. The vertical dotted line shows the frequency corresponding to the 108\,d period.}
\label{four}
\end{figure}

Finally, we performed period searches on the line profiles using the modified Fourier algorithm \citep{hmm} and the same intervals as for the moment calculation. To this aim, we interpolated all profiles to a common velocity grid with steps of 0.2\,\kms, and then calculated periodograms at each velocity considering the same velocity intervals as used for moment calculations. These individual periodograms were then averaged to get a single periodogram per line (Fig. \ref{four}). While no frequency stands out, multiple peaks often appear near 0.015-0.025\,d$^{-1}$ for absorption lines, hinting at long-term variability on timescales of 40--65 days, that is, close to about half the X-ray period. Indeed, spectra taken only a few days apart (2017 May 24 and 29) show no significant difference while variations are detected on longer timescales, typically tens of days (see Figs. \ref{photom} and \ref{rvphot} - e.g. in the latter figure, the times of the two minimum velocities are separated by $\sim$50\,d).

\section{Discussion}
The previous sections led to several results. Firstly, at high energies, the flux varies by a factor of about two and a period of 108\,d is detected for these changes. The minimum flux corresponds to a time of lower intrinsic emission and increased absorption. Secondly, we confirmed the presence of photometric and spectroscopic changes in the star, which we further constrained. However, no regular clock seems present in the optical/near-IR domain, a simple timescale of 50--100\,d being found. We now attempt to interpret these contrasting features.

\subsection{The evolutionary status of \duz}

The second data release of $GAIA$ \citep{Brown} allowed for the first time a direct measurement of the parallax of \duz\ to be made. This led to a surprising value of 1.18$\pm$0.13\,mas. Indeed, this must be compared to the DR2 parallaxes of the other three main objects of the Cyg\,OB2 association, which are compatible with the usual distance used for Cyg\,OB2 \citep[e.g. $d\sim$1.75\,kpc in][]{cla12}: 0.64$\pm$0.06\,mas for Cyg\,OB2\,\#5, 0.62$\pm$0.04\,mas for Cyg\,OB2\,\#8A, and 0.60$\pm$0.03\,mas for Cyg\,OB2\,\#9. \duz\ thus appears twice closer than the stars with which it is often associated. Proper motions also appear quite different, with values of $-1.9\pm0.2$\,mas in RA and $-3.3\pm0.2$\,mas in DEC for \duz\ and values of minus 2.7--3.1\,mas in RA and minus 4.1--4.7\,mas in DEC for the others. This actually confirms early doubts on the membership of \duz\ to Cyg\,OB2 \citep{wal73}.

One may however wonder whether or not the presence of visual companions to \duz\ could impair the $GAIA$ measurements. In this context, it is important to recall that all four X-ray bright massive stars of Cyg\,OB2  are actually multiple, with very diverse properties: Cyg\,OB2\,\#8A and 9 consist of spectroscopic binaries with  periods of 22\,d and 2.4\,yr, respectively, while Cyg\,OB2\,\#5 is thought to have four components with periods ranging from 6.6\,d to 9000\,yrs. If companions were modifying the parallax and proper motions of \duz, it would be very surprising to find that the data of the other three systems agree - but they do. Therefore the parallax difference of \duz\ is very probably real. In this context, one may also argue that the angular diameter of \duz\ amounts to 1.31 mas, i.e.\ comparable to the parallax, as is the case of other blue supergiants ($\zeta^1$\,Sco, HD\,190603, BP\,Cru, HD\,80077, HD\,168607, and to a lesser extent P\,Cygni and HD\,168625). Shifts in the photocentre of the star (e.g.\ due to non-radial pulsations or spots on the stellar surface; see below) could thus affect the parallax. However, we note that the absence of a strong, strictly periodic photometric modulation does not support this scenario. Furthermore, in view of the rather small relative error on the {\it GAIA}-DR2 parallax, it would require an extraordinary coincidence for the shifts of the photocentre to exactly mimic the effect of an annual parallax in the {\it GAIA} data. We thus consider the revision of the distance as robust.

This new distance ($840^{945}_{755}$\,pc in \citealt{bai18}) drastically changes the accepted physical properties of \duz: scaling the values of \citet{cla12} yields a luminosity of $\log[L_{\rm BOL}/L_{\odot}]=5.64$, a radius of 118\,R$_{\odot}$, and a spectroscopic mass of 25\,M$_{\odot}$. The mass-loss rate proposed by \citet{cla12} or \citet{mor16} would also be reduced by about a factor of four. \duz\ now appears as a normal supergiant, without any luminosity problem \citep[a difficulty noted by ][]{cla12}.

It is interesting to compare \duz\ with the LBV R~71 in the Large Magellanic Cloud which is at the lower luminosity end of the classical LBV strip \citep{Mehn17}. Both stars share similar properties \citep[and this work]{Mehn17,cla12}: they have similar spectroscopic masses (27\,M$_{\odot}$ for R~71 and 26\,M$_{\odot}$ for \duz), temperatures (15\,500$\pm$500\,K for R~71 in quiescence and $13\,700^{+800}_{-500}$\,K for \duz), radii (107\,R$_{\odot}$ for R~71 in quiescence and 118\,R$_{\odot}$ for \duz), and luminosities ($6.0\,10^5$\,L$_{\odot}$ for R~71 in quiescence and $4.4\,10^5$\,L$_{\odot}$ for \duz). The slight luminosity difference is sufficient to keep \duz\ outside of the LBV strip, but actually close to it, while R~71 lies inside of it. Evidence that \duz\ is less evolved than R~71 comes from the chemical composition of its atmosphere. \citet{Mehn17} reported a strong enrichment of the atmosphere of R~71 with CNO processed material (with number ratios of He/H = 0.20, O/N = 0.08 and C/N = 0.05). For \duz, \citet{cla12} derived a significantly weaker enrichment (He/H = 0.1, O/N = 1.48, C/N = 0.31). 

\begin{figure}
\begin{center}
\includegraphics[width=8.5cm]{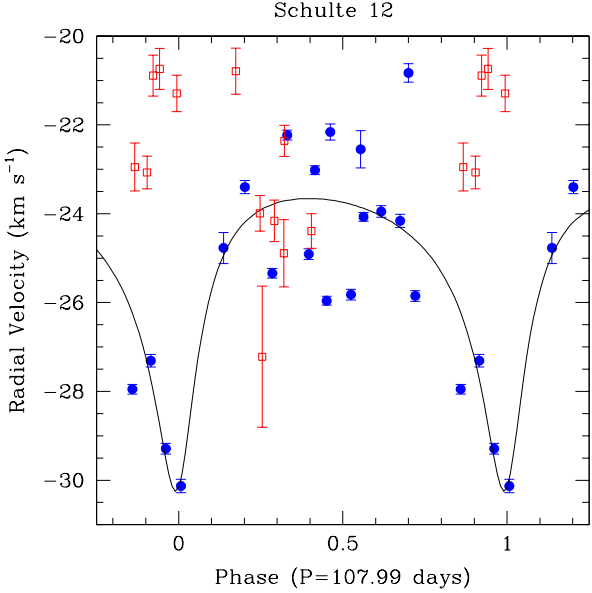}
\end{center}  
\caption{Tentative orbital solution of \duz\ for a fixed orbital period of 108\,d (middle column of Table\,\ref{solorb}). The average velocities of Mg\,{\sc ii} and Fe\,{\sc ii} emissions in the Carmenes dataset are shown by black dots and the Hermes velocities of Fe\,{\sc ii}\,$\lambda$\,7513\AA\ (not used to obtain the orbital solution) by open red squares. We note that the phase used here corresponds to the binary ephemeris (Table \ref{solorb}), not to the X-ray ephemeris used for Figs. \ref{lc2} and \ref{figfit}.}
\label{RVcurve}
\end{figure}

\subsection{Binarity}
As already mentioned in Section 4.2, the changing profile of the absorption lines cannot be due to line blending as the profile broadens, skews, or becomes narrower without much changes in line depth (Fig. \ref{spec}). However, the Mg\,{\sc ii} and Fe\,{\sc ii} emissions undergo simple velocity shifts, which could be interpreted in this framework. Combining the centroid velocities of the Mg\,{\sc ii} and Fe\,{\sc ii} emission lines, we computed using LOSP\footnote{LOSP \citep{san06} is an improved version of the code originally proposed by \citet{wol67}.} a highly preliminary orbital solution. In this calculation, we assume that the 108\,d X-ray period is linked to orbital motion. Freeing the period does not significantly change the quality of the fit or the values of the orbital parameters. The results are given in Table\,\ref{solorb} and illustrated in Fig.\,\ref{RVcurve}. The periastron passage of this tentative orbit occurs $\sim$10\,d (or $\Delta(\phi)=$0.1) after the X-ray maximum and $\Delta(\phi)=$0.25 before the X-ray minimum - in colliding-wind binaries however the periastron usually corresponds quite closely to an X-ray extremum.

\begin{table}
\caption{Tentative orbital solution built from the mean RVs of the Mg\,{\sc ii}\,$\lambda$\,10952\AA\ and Fe\,{\sc ii}\,$\lambda\lambda$\,7513,\,9998\AA\ emission lines (Carmenes data only). \label{solorb}}
\setlength{\tabcolsep}{3pt}
\begin{tabular}{l c c}
  \hline
$P_{\rm orb}$ (days)     & 108.0 (fixed)  & $104.1 \pm 5.7$ \\      
$\gamma$ (km\,s$^{-1}$) & $-25.2 \pm 0.5$ & $-25.2 \pm 0.5$ \\
$K$ (km\,s$^{-1}$)      & $3.3 \pm 0.8$   & $3.3 \pm 0.8$ \\
$e$                    & $0.55 \pm 0.20$ & $0.53 \pm 0.20$\\
$\omega$ ($^{\circ}$)   & $193 \pm 23$    & $197 \pm 25$ \\
$T_0$ (HJD$-$2\,450\,000) & $8299.8 \pm 5.4$ & $8300.4 \pm 5.9$ \\
$a\,\sin{i}$ (R$_{\odot}$) & $5.9 \pm 1.7$ & $5.8 \pm 1.6$ \\
$f(m)$ (M$_{\odot}$)    & $0.00024 \pm 0.00020$ & $0.00024 \pm 0.00046$ \\
rms (km\,s$^{-1}$)      & $1.7$ & $1.7$ \\
\hline
\end{tabular}
\end{table}

If we assume \duz\ to be a binary system, we need to ask ourselves whether or not the spectroscopic radius of the B supergiant is consistent with the radius of its Roche lobe. For this purpose, we use the formula of \citet{Eggleton} to express the dependence of the Roche lobe radius on the mass-ratio $q = \frac{m_1}{m_2}$ as follows:
\begin{equation}
\frac{R_{\rm RL}}{R_{\odot}} = \frac{2.06\,q^{2/3}}{0.6\,q^{2/3}+\ln{(1 + q^{1/3})}}\,\left(\frac{1+q}{q}\right)^{1/3}\,\left(\frac{m_1}{M_{\odot}}\right)^{1/3}\,\left(\frac{P}{day}\right)^{2/3}
.\end{equation}
If we adopt a period of 108\,d and a spectroscopic mass of $m_1 = 25$\,M$_{\odot}$ for the B supergiant, we find that the radius of the Roche lobe of the  B supergiant exceeds 127\,R$_{\odot}$ whatever the value of $q$. Therefore, if the orbit were circular, the spectroscopic radius of 118\,R$_{\odot}$ would fit entirely into the Roche lobe. The B supergiant would fill less than 80\% of the volume of its Roche lobe, regardless of the mass of its companion. However, this is valid for a circular orbit, not in the presence of a significant orbital eccentricity as found here. Assuming that in a long-period eccentric binary the size of the Roche lobe varies in proportion to the instantaneous orbital separation, we find that for an eccentricity of about 0.5 (Table \ref{solorb}) the B supergiant would significantly overflow its Roche lobe at phases near periastron, regardless of the value of the mass of the companion. 

Nevertheless, there is a significant dispersion of the data points around the best-fit curve, and moreover, plotting the additional Fe\,{\sc ii}\,$\lambda$\,7513\AA\ velocities measured on Hermes spectra does not reveal a coherent behaviour (Fig. \ref{RVcurve}). It must be underlined here once again that Hermes and Carmenes spectra agree well: the spectra taken in May 2017 with both instruments at only a 5\,d time difference appear very similar while the DIB centroids coincide within errors (see Table \ref{averm}). Therefore, no problem in the instrument or during the analysis can be blamed for the difference observed in Fig. \ref{RVcurve}. In fact, no orbital solution using the full Fe\,{\sc ii}\,$\lambda$\,7513\AA\ velocity dataset can be derived as these data simply cannot be reconciled with a 108\,d period. Combined to the partial correlation of the emission line velocities with those of absorptions (Figs. \ref{correlm} and \ref{rvphot}), this suggests that we are not witnessing pure orbital motion. If \duz\ is a binary, we have not yet found evidence for it in the RVs. 

Finally, we examined the spectrum of \duz\ in search of lines moving in the opposite direction compared to Mg\,{\sc ii} or of lines with a different ionization (e.g. He\,{\sc ii}), as they could then be attributed to a companion. None were found however. With the revised distance of \duz, its absolute $V$-band magnitude amounts to $-8.26$. We estimate that our spectroscopic data should enable us to detect the spectroscopic signature of an O-type companion down to 2.5\,magnitudes fainter than the B supergiant. Considering the typical parameters of O-type stars from the calibration of \citet{Martins}, we conclude that a main-sequence O-type star would remain entirely hidden, regardless of its spectral type. For O-type giants, only objects with a spectral type earlier than O5 would show up in the combined spectrum. Finally, O supergiants should be detectable in a combined spectrum regardless of the spectral types. The non-detection of even a massive companion would therefore not be entirely surprising.

\begin{figure}
\includegraphics[width=8.5cm]{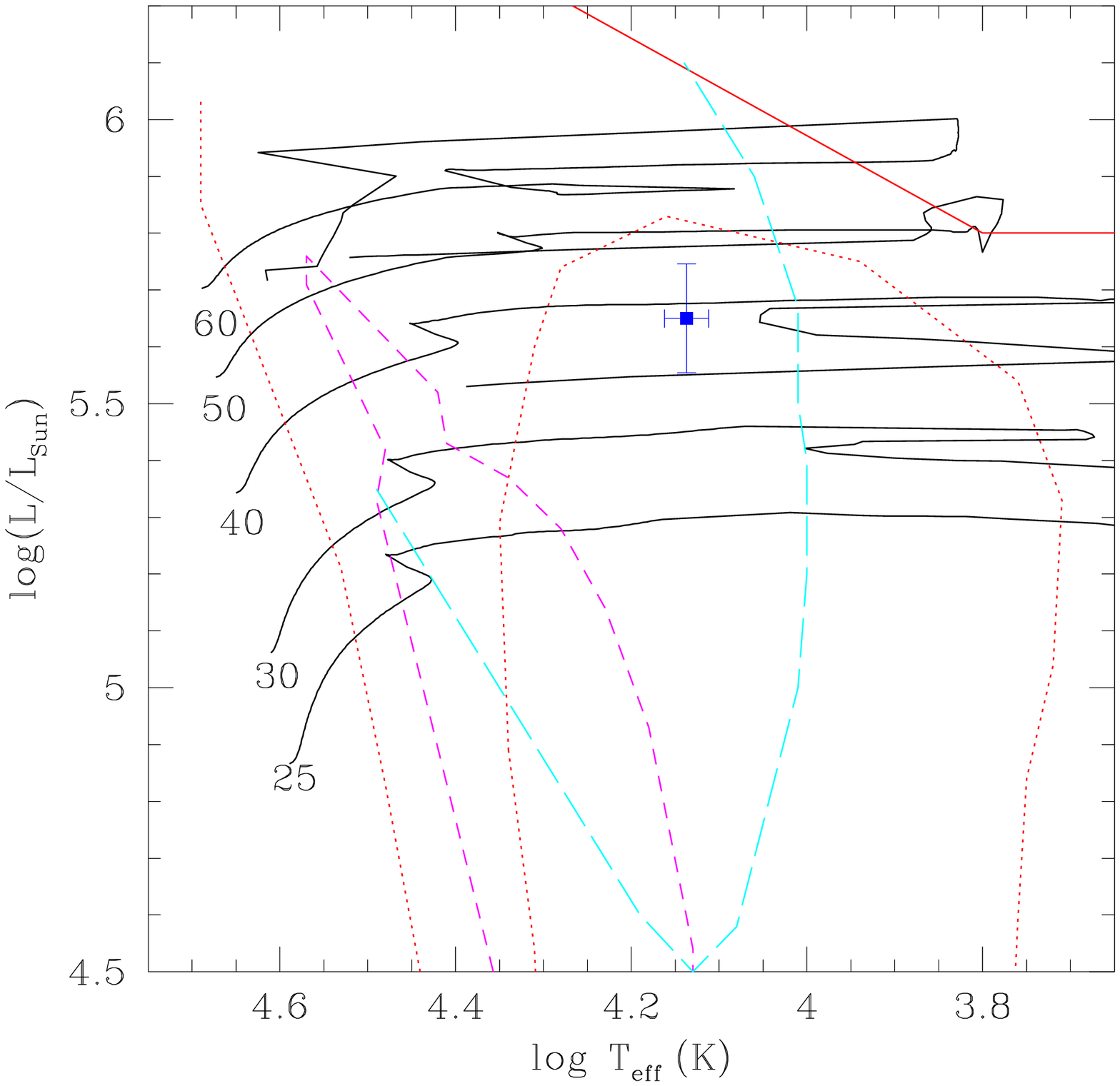}
\caption{Revised position of \duz\ in the Hertzsprung-Russell diagram adopting the distance corresponding to the {\it GAIA}-DR2 parallax. The evolutionary tracks are taken from \citet{Ekstrom12} for solar metallicity and no rotation. The continuous red line corresponds to the empirical Humphreys-Davidson limit \citep{HD}. Various instability boundaries are shown according to \citet{Saio}. The dotted red curve indicates the instability boundary of low-order radial and non-radial p-mode oscillations. The short-dashed magenta contour indicates the region where non-radial g-modes with ${\ell} \leq 2$ are excited by the iron opacity bump. Finally, the cyan long-dashed contour indicates the region where oscillatory convection modes with ${\ell} \leq 2$ are excited \citep{Saio}.}
\label{HRDGeneva}
\end{figure}

\subsection{Pulsational activity}
The erratic variations of the centroids and profiles of the absorption lines suggest that the velocities of these lines are dominated by effects other than genuine orbital motion. One explanation for these variations could be non-radial pulsations of the B supergiant; but is \duz\ expected to be pulsationally unstable?

With the revised physical properties, the position of \duz\ in the Hertzsprung-Russell diagram (Fig.\,\ref{HRDGeneva}) drastically changes. The star now appears relatively far away from the empirical Humphreys-Davidson limit \citep{HD} and thus from the region where {\it strange-mode oscillations} are expected \citep{Kiriakidis}. \citet{Maeder} suggested that intermediate term variability of B -- G supergiants could stem from non-radial gravity-mode pulsations connected to a large outer convection zone. Indeed, \duz\ is located within the part of the Hertzsprung-Russell diagram where oscillatory convection modes with ${\ell} \leq 2$ are expected \citep{Saio}. If we attribute the photometric variability of \duz\ to non-radial pulsations, they must indeed be of low ${\ell}$ to explain the relatively large observed amplitude of the variations. For a 30\,M$_{\odot}$ star, \citet{Saio} predicts periods of excited ${\ell} = 1$ and ${\ell} = 2$ modes of about 70 and 60\,days, respectively. These values are reasonably close to the timescale of the variations of the absorption lines and the photometry of \duz. Furthermore, the pulsation modes are associated to subphotospheric convection which is known to be unstable, hence the pulsational activity could be intermittent, and they have growth timescales comparable to their period, suggesting that such (dis)appearance could occur quite quickly, in line with our observations of \duz.

The line profile variability (LPV) of absorption lines due to non-radial pulsations depends on the ratio $k$ between the azimuthal and radial pulsation velocities \citep{Kambe}. The fact that the strongest LPV signatures are observed in the cores of the lines (see Fig. \ref{tvs}), suggests that $k << 1$. \duz\ lies also at the edge of the zone where low-order ($n = 0$ or $n=1$) radial pressure modes are expected to be excited according to \citet{JS16}. However, the corresponding fundamental ($n = 0$) radial modes should have periods of order $10^6$\,s, that is\ about 12\,days, which is much shorter than the observed periods. It seems therefore unlikely that the observed variability of \duz\ could arise from such radial modes.

At this stage, it is once more interesting to compare \duz\ with the LBV R~71. During quiescence, R~71 displays microvariations of 0.1\,mag amplitude on timescales of 14 -- 100\,days \citep[][and references therein]{Mehn17}. This is qualitatively similar to what we observe in \duz, probably indicating that both stars are subject to similar pulsation mechanisms. During the currently ongoing outburst of R~71, a new kind of nearly periodic photometric variation appeared, consisting of oscillations of 0.2\,mag amplitude on a timescale of about 445\,days \citep{Wal17,Mehn17}. The star R~71 currently displays a late F supergiant spectrum \citep{Wal17,Mehn17}, very different from that of \duz. However, an interesting point concerns the spectroscopic variability that occurs on a timescale twice that of the above-mentioned 445-day photometric cycle. Indeed, the absorption lines vary between single and double morphologies \citep{Mehn17,Wal17}, in a way similar to what we observe for \duz. By analogy with RV~Tau variables, \citet{Mehn17} attribute the line doubling to shock waves triggered by the pulsations and propagating through the atmosphere of R~71. Whether a similar explanation holds in our case is unclear, as the atmospheric structure of the B5\,Ia supergiant \duz\ is rather different from that of a late-F supergiant.

\subsection{The X-ray clock}
The X-ray data for \duz\ display a clear periodicity of 108\,d. This is incompatible with the putative periods of the visual companions (see Introduction), and therefore they do not play a role in this phenomenon. The question then arises regarding the origin of the high-energy emission.

The folded X-ray light curve of \duz\ is strongly reminiscent of that of the massive binary WR\,21a, where X-rays arise in the collision between the stellar winds of the two stars \citep{gos16}. As in \duz, the count rate remains relatively constant most of the time (or slightly increases), then a stronger increase occurs, followed by a sharp drop and a longer recovery. In WR\,21a, the increase was associated to the typical $1/D$ increase in flux found in adiabatic wind--wind collisions as stars approach periastron \citep{ste92}. The subsequent drop combined two effects: an initial increase of absorption as the companion plunged towards the densest wind layers at periastron, then a shock collapse followed by a slow recovery \citep{gos16}. The two X-ray light curves thus appear very similar, except for the amplitude of the X-ray variation (a factor of five between minimum and maximum in WR21a rather than two in \duz) but this can easily be accounted for by different physical and orbital parameters, as the X-ray properties are very sensitive to them. Further supporting this scenario is the presence of very hot plasma (demonstrated by the need of $kT_2$=1.9\,keV and the presence of the iron line at 6.7\,keV), as well as the strong X-ray luminosity of \duz\ (the flux in the 0.5--10.\,keV energy band, corrected for interstellar absorption, lies between 5.6 and 11.9$\times 10^{-12}$\,erg\,cm$^{-2}$\,s$^{-1}$, corresponding to 4.7--10.$\times 10^{32}$\,erg\,s$^{-1}$ or $\log[L_{\rm X}/L_{\rm BOL}]=-6.6$ to --6.2). Both are common features in wind--wind collisions.

However, to produce a significant wind--wind collision, the companion of \duz\ ought to be a massive star of spectral-type O, early B, or Wolf-Rayet with a fast wind and in an eccentric orbit (to produce the $1/D$ effect). From the spectral analysis in the optical, such a companion cannot be an O-type supergiant or early giant, but other possibilities remain open in view of the brightness of \duz. Nevertheless, as the spectroscopic mass of \duz\ is rather modest, the required massive OBWR companion would then lead to significant orbital motion for \duz, except if the orbital inclination is low. In that case however an increase of the absorbing column towards a putative wind--wind collision zone could only occur at periastron, contrary to observations. In fact, as mentioned above, we certainly cannot claim to have detected the binarity nature of \duz, hence the wind--wind collision scenario appears quite unlikely.

As the optical analysis showed, non-radial pulsations are most probably at work in \duz. Could the X-ray light curve then be driven by pulsation? X-ray pulsations occurring in phase with the optical pulsations have indeed been observed for a few $\beta$\,Cep-type pulsators \citep{NatCom,xiCMa,caz14}. There are however a number of caveats with this hypothesis. Firstly, the X-ray light curves of $\xi^1$\,CMa \citep{NatCom,xiCMa} and HD\,44743 \citep{caz14} are nearly sine waves, whereas the X-ray light curve of \duz\ is far more asymmetric. It must also be mentioned that the (X-ray) pulsation periods of the two $\beta$\,Cep stars are much shorter (4.9 and 6.0\,h for $\xi^1$\,CMa and HD\,44743, respectively) than the period of the X-ray light curve of \duz. Secondly, the peak-to-peak amplitudes of the modulation of the X-ray fluxes in $\xi^1$\,CMa and HD\,44743 are less than 10\% of the mean flux. In the case of \duz\, this relative peak-to-peak amplitude reaches 70\% of the mean flux. Thirdly, the modulation of the X-ray fluxes of the two $\beta$\,Cep stars occurs exactly on the same period as the modulation of their optical light curve. In the case of \duz, the X-ray flux is modulated on a period roughly twice that of the optical variations (when they exist). Lastly, with $\log[{L_{\rm X}/L_{\rm bol}}] = -6.7$ and $-7.6$ for $\xi^1$\,CMa and HD\,44743, respectively, the level of X-ray emission of the two $\beta$\,Cep stars is very close to (or slightly below) the typical one of massive stars, while \duz\ clearly appears overluminous in X-rays. The two $\beta$\,Cep stars also display the typical soft X-ray emission ($kT<1$\,keV, \citealt{naz14,caz14}). There is therefore no observational evidence in favour of this scenario, and a detailed modelling for the peculiar pulsations of \duz\ would be required to demonstrate their influence on the X-ray emission. 

Another candidate for a regular clock is rotation, as the 108\,d period is compatible with the rotation estimates of \duz. \citet{cla12} derived $v\,\sin{i} = 38$\,km\,s$^{-1}$ for \duz: adopting a radius of 118\,R$_{\odot}$ leads to an upper limit on the rotational period of 158\,days. Likewise, adopting the spectroscopic mass, radius, and luminosity, we infer a lower limit on the rotational period of 41\,days, assuming the equatorial rotation velocity reaches the break-up velocity. Any observed periodicity in this rather wide range could thus possibly stem from rotational modulation. Two possible scenarios can be envisaged in this context: spots on the stellar surface leading to co-rotating interaction regions \citep[CIRs,][]{Cranmer,Lobel}, and a magnetically confined stellar wind \citep{Babel,ud-Doula}.

Considering the first case, it should be noted that variability of the H$\alpha$ emission line is ubiquitous among OBA-type supergiants such as \duz\ \citep[e.g.][]{Fullerton,Kaufer99,Morel2004}. Based on the similarity between the observed recurrence timescales and the estimated rotation periods (or an integer fraction of them), this phenomenon is frequently attributed to co-rotating large-scale structures in the wind \citep{Fullerton}. Investigating a sample of 22 OB supergiants with spectral types in the range O7.5 to B5, where \duz\ lies, \citet{Morel2004} found significant H$\alpha$ variability for all of them. These latter authors further reported the existence of a cyclical pattern of variability in the H$\alpha$ profile of nine stars of their sample. They also noted the existence of periodicities ranging between 1 and 27\,days in the photometric data of nine stars. Hence the presence of CIRs in \duz\ would not be a surprising occurrence. In this context, several cases of rotational modulation of the X-ray flux of O-stars have been reported in recent years and were attributed to such large-scale structures \citep{osk01,naz13,xiPer,lamCep,zetaPup}. However, their variation timescales are shorter (1.8--4.2\,d) than for \duz\ and the peak-to-peak amplitude of their modulations is $\sim$10\% of the mean flux, which is much smaller than for \duz. Furthermore, the X-ray brightness of these stars are in agreement with expectations for massive stars and their plasma temperature remains low, $kT<1$\,keV \citep{naz09,lamCep,nazoe}, contrasting with \duz. Finally, while the optical light curve of \duz\ is strongly epoch-dependent, the photometric variations of such stars remain relatively coherent over timescales of years (see \citealt{Morel2004} for a sample of B-supergiants and \citealt{ram14,ram18} for $\xi$\,Per and $\zeta$\,Pup, respectively). There is thus poor observational support for this scenario.

In the second case, X-ray overluminosities, large X-ray modulations, and large plasma temperatures may occur \citep{naz14} but this observational picture was mostly drawn from magnetic main sequence stars and therefore its applicability to supergiants remains to be demonstrated. Magnetic fields are seldom detected in massive supergiants, not only because of a (expected) low incidence rate but also for technical reasons. Indeed, as the stars evolve, their radii increase and their magnetic fields, if descending from the main-sequence fields, should be much smaller and therefore more difficult to detect. For these reasons, only a few magnetic detections of supergiants have been achieved until now \citep{bla15,mar18} and for $\zeta$\,Ori\,A (O9.5I) the X-ray emission appears in line with the `typical' properties of massive stars, that is it seems unaffected by the presence of the magnetic field \citep{naz14}. The applicability of this scenario is therefore suspended to a detailed investigation of the wind properties of \duz\ combined with the (positive) results of a spectropolarimetric campaign. In addition, our optical spectroscopy did not reveal a strict periodic modulation of the H$\alpha$ emission line contrary to observations and expectations of magnetically confined winds \citep{tow05,sun12,udd13}.

A last, tentative possibility would be sporadic accretion onto a companion of the B-star in an eccentric orbit. The pulsations could then be unrelated or excited through tidal interaction at periastron. However, the X-ray emission of \duz\ is thermal in nature, thus different from the typical non-thermal emission associated to accretion (though some peculiar accretion regimes may produce thermal emission). Furthermore, the X-ray variations concern not only the intrinsic strength of the emission but also the local absorption and therefore they are not in line with a simple additional emission appearing and disappearing at some phases.

\section{Conclusions}
We performed a multiwavelength campaign dedicated to \duz, which we complemented with archival data. This unique dataset sheds new light on the properties of this star.

Firstly, an X-ray clock is present in the system: a period of 108\,d, with a minimum flux occurring $\sim$35\,d later than the maximum flux. This minimum corresponds to an increase in absorption along the line of sight and a decrease in the intrinsic emission of \duz. The same phased light curve shape is found (1) from data scattered over a 14-year timescale (or 47 cycles, demonstrating the stability of the signal) as well as in a dedicated monitoring covering a full cycle (enabling us to reject aliases or incorrect period identifications); and (2) from fully independent instruments onboard the \sw\ and \xmm\ telescopes, which observed \duz\ with very different temporal sampling. The period is detected by different period-search methods dedicated to uneven samplings; its significance level (modified Fourier, AOV) is $\le$0.02\%. Therefore the periodicity does not appear to be spurious. 

Secondly, we confirmed the spectral type of \duz\ as well as previous reports of its photometric and spectroscopic variability, which we further characterised. While not strictly periodic, their timescale amounts to tens of days, in agreement with the X-ray period (or rather half of it). In the visible spectrum, all absorption lines of hydrogen, helium, and metals behave in a similar way. Their strengths (EWs) and widths do not vary much, but their centroids and shapes suffer from significant changes - for example velocities cover a range of about 25\,\kms, while lines alternate between blueward-skewed, rightward-skewed, symmetric, and doubling profiles. The velocities of H and He\,{\sc i} lines appear to lag by a few days behind those of metals, probably due to different formation depths. Emission lines of Mg\,{\sc ii} and Fe\,{\sc ii} appear less variable, with profiles simply shifting by 10\,\kms, but these variations, which are larger than the noise level, do not seem to indicate binary motion. Finally, H$\alpha$ and He\,{\sc i}\,$\lambda$\,10830\AA, the two lines most affected by wind emission, display spectacular changes, of an amplitude larger than reported before. The behaviour of these lines does not seem obviously linked to that of the other lines and they do not exhibit strict periodicities either. 

The {\it GAIA} data reveal that \duz\  is\ actually twice closer than other massive stars in Cyg\,OB2. This led to a severe revision of its stellar parameters: \duz\ is no longer a hypergiant, but a normal supergiant, and is no longer close to the empirical Humphreys-Davidson instability limit; it is however located in an instability zone linked to convection modes for which periods of tens of days are expected. This certainly explains the changes in the absorption line profiles, though the absence of strict periodicity implies intermittent activation of the oscillations.

Regular modulations of the X-ray emission of massive stars have been observed in several cases, but none seem to strictly apply to \duz. The amplitude and shape of the X-ray light curve, the presence of a hot plasma, and the significant X-ray overluminosity recall the characteristics of colliding-wind binaries. However, the lack of any manifest binary signature in the optical/near-IR spectrum seems to exclude this possibility. In a few cases, pulsations have been linked to X-ray activity, but without overluminosity or high-temperature plasma and with shorter timescales and smaller variability amplitude than observed for \duz. Rotational modulations of the X-ray light curves also exist. When attributed to CIRs, they again lack the overluminosity, hot plasma, and large variability amplitude observed in \duz. When linked to the presence of magnetic field, such properties are encountered, but only for main sequence stars. The origin of the X-ray clock therefore remains mysterious. Further investigation, for example of UV wind lines, is now required.

\begin{appendix}
  \section{Complementary table and images}
\begin{table*}
\centering
\footnotesize
\caption{Journal of the X-ray observations. HJD correspond to dates at mid-exposure.}
\label{journal}
\begin{tabular}{lcccc}
\hline\hline
\multicolumn{5}{l}{\xmm\ - Count rates in total band ($HR=(H-S)/(H+S)$) for MOS1, MOS2, and pn}\\
ObsID & HJD-2\,450\,000. & \multicolumn{3}{c}{Count Rates (cts\,s$^{-1}$)}\\
\hline
0896/0200450201 & 3308.581 & 0.287$\pm$0.006 (--0.13$\pm$0.02) &0.294$\pm$0.006 (--0.10$\pm$0.02) & 0.747$\pm$0.023 (--0.230$\pm$0.030) \\
0901/0200450301 & 3318.559 & 0.310$\pm$0.005 (--0.12$\pm$0.02) &0.326$\pm$0.006 (--0.08$\pm$0.02) & 0.799$\pm$0.010 (--0.232$\pm$0.012) \\
0906/0200450401 & 3328.543 & 0.352$\pm$0.006 (--0.15$\pm$0.02) &0.371$\pm$0.006 (--0.11$\pm$0.02) & 0.887$\pm$0.010 (--0.227$\pm$0.010) \\
0911/0200450501 & 3338.505 & 0.314$\pm$0.007 (--0.24$\pm$0.02) &0.339$\pm$0.008 (--0.21$\pm$0.02) & 0.775$\pm$0.012 (--0.318$\pm$0.014) \\
1353/0505110301 & 4220.354 & 0.208$\pm$0.007 (--0.16$\pm$0.03) &0.202$\pm$0.007 (--0.15$\pm$0.03) & 0.551$\pm$0.014 (--0.191$\pm$0.025) \\
1355/0505110401 & 4224.169 & 0.219$\pm$0.006 (--0.19$\pm$0.03) &0.202$\pm$0.006 (--0.12$\pm$0.03) & 0.577$\pm$0.012 (--0.200$\pm$0.021) \\
2114/0677980601 & 5738.256 &                                   &                                  & 0.510$\pm$0.008 (--0.261$\pm$0.015) \\
2625/0740300101 & 6758.209 & 0.230$\pm$0.006 (--0.14$\pm$0.03) &0.247$\pm$0.005 (--0.19$\pm$0.02) & 0.648$\pm$0.010 (--0.305$\pm$0.015) \\
3089/0780040101 & 7683.229 & 0.199$\pm$0.005 (--0.20$\pm$0.02) &0.198$\pm$0.005 (--0.19$\pm$0.02) & 0.506$\pm$0.010 (--0.239$\pm$0.022) \\
3097/0793183001 & 7699.380 &                                   &0.325$\pm$0.008   (0.14$\pm$0.02) &                                     \\
3176/0800150101 & 7856.833 & 0.276$\pm$0.006 (--0.19$\pm$0.02) &0.274$\pm$0.007 (--0.15$\pm$0.02) & 0.730$\pm$0.012 (--0.288$\pm$0.016) \\ 
3273/0801910201 & 8050.426 &                                   & 0.336$\pm$0.010 (0.16$\pm$0.03) &                                      \\
3280/0801910301 & 8063.415 &                                   & 0.340$\pm$0.012 (0.11$\pm$0.03) &                                      \\
3284/0801910401 & 8071.063 &                                   & 0.385$\pm$0.017 (0.15$\pm$0.04) &                                      \\
3288/0801910501 & 8079.696 &                                   & 0.370$\pm$0.012 (0.08$\pm$0.03) &                                      \\
3294/0801910601 & 8092.351 &                                   & 0.407$\pm$0.016 (0.00$\pm$0.04) &                                      \\
\hline
\multicolumn{5}{l}{\sw\ - Count rates in total band ($HR=H/S$), UVOT magnitudes}\\
ObsID & HJD-2\,450\,000. & Count Rates ($HR=H/S$) & UVOT \\
      &                  & (cts\,s$^{-1}$)         & (mag) \\
\hline
00037920001 & 4657.335 & 0.038$\pm$0.006 (0.82$\pm$0.25) & $^a$\\
00031904001 & 5571.617 & 0.069$\pm$0.004 (0.73$\pm$0.09) & $UVW1$=16.85$\pm$0.04\\
00031904002 & 5655.834 & 0.055$\pm$0.004 (0.70$\pm$0.10) & $UVW1$=16.86$\pm$0.04\\
00031904003 & 5700.082 & 0.092$\pm$0.005 (0.63$\pm$0.08) & \\
00031904004 & 5743.841 & 0.061$\pm$0.004 (0.65$\pm$0.08) & $UVW1$=16.88$\pm$0.04\\
00031904005 & 5842.171 & 0.046$\pm$0.004 (0.70$\pm$0.11) & $UVM2$=21.06$\pm$0.55, $UVW2$=17.81$\pm$0.05\\
00032767001 & 6380.437 & 0.043$\pm$0.005 (0.64$\pm$0.14) & $UVW1$=16.91$\pm$0.05\\
00032767002 & 6380.874 & 0.050$\pm$0.004 (0.66$\pm$0.11) & $U$=16.13$\pm$0.03\\
00033818001 & 7191.528 & 0.069$\pm$0.012 (0.91$\pm$0.32) & $UVW1$=16.87$\pm$0.05\\
00033818002 & 7192.049 & 0.069$\pm$0.036 (0.67$\pm$0.61) & $UVW1$=16.86$\pm$0.11\\
00033818003 & 7192.738 & 0.069$\pm$0.012 (0.87$\pm$0.31) & $U$=16.24$\pm$0.04\\
00033818005 & 7193.317 & 0.053$\pm$0.010 (0.81$\pm$0.29) & \\
00033818004 & 7193.513 & 0.090$\pm$0.021 (0.90$\pm$0.43) & \\
00033818006 & 7194.261 & 0.075$\pm$0.010 (0.63$\pm$0.18) & \\
00033818007 & 7194.532 & 0.055$\pm$0.010 (0.64$\pm$0.24) & \\
00033818008 & 7195.319 & 0.071$\pm$0.010 (0.64$\pm$0.19) & \\
00033818009 & 7195.784 & 0.067$\pm$0.008 (0.95$\pm$0.23) & $UVW1$=16.87$\pm$0.05\\
00033818010 & 7196.160 & 0.082$\pm$0.011 (0.87$\pm$0.24) & $UVW1$=17.02$\pm$0.08\\
00033818011 & 7196.661 & 0.073$\pm$0.011 (0.58$\pm$0.18) & $U$=16.28$\pm$0.03\\
00033818012 & 7197.114 & 0.067$\pm$0.013 (0.53$\pm$0.22) & \\
00033818013 & 7197.917 & 0.079$\pm$0.010 (0.54$\pm$0.15) & \\
00033818014 & 7198.438 & 0.089$\pm$0.019 (0.46$\pm$0.23) & \\
00033818016 & 7199.052 & 0.085$\pm$0.012 (1.18$\pm$0.34) & \\
00033818017 & 7199.512 & 0.052$\pm$0.013 (0.57$\pm$0.31) & $UVW1$=16.81$\pm$0.06\\
00033818018 & 7200.299 & 0.065$\pm$0.012 (0.75$\pm$0.27) & \\
00033818020 & 7201.233 & 0.050$\pm$0.011 (1.10$\pm$0.47) & $U$=16.30$\pm$0.03\\
00033818021 & 7201.581 & 0.107$\pm$0.013 (1.00$\pm$0.25) & \\
00033818022 & 7202.031 & 0.079$\pm$0.010 (0.69$\pm$0.18) & $UVW2$=17.46$\pm$0.06\\
00033818023 & 7202.535 & 0.066$\pm$0.011 (0.62$\pm$0.22) & $UVM2$=21.77$\pm$1.42\\
00033818024 & 7203.511 & 0.096$\pm$0.017 (0.55$\pm$0.20) & \\
00033818025 & 7204.158 & 0.096$\pm$0.013 (1.07$\pm$0.29) & $UVW1$=16.87$\pm$0.05\\
00033818026 & 7204.891 & 0.064$\pm$0.014 (0.71$\pm$0.31) & $U$=16.32$\pm$0.03\\
00033818027 & 7205.091 & 0.071$\pm$0.010 (0.59$\pm$0.17) & \\
00033818028 & 7205.757 & 0.079$\pm$0.010 (0.64$\pm$0.17) & $UVW2$=21.92$\pm$1.08\\
00033818029 & 7206.101 & 0.089$\pm$0.012 (0.55$\pm$0.15) & \\
00033818031 & 7207.032 & 0.087$\pm$0.012 (0.60$\pm$0.17) & \\
00033818032 & 7207.752 & 0.069$\pm$0.010 (0.67$\pm$0.20) & $UVW1$=16.85$\pm$0.05\\
00033818033 & 7208.098 & 0.073$\pm$0.010 (1.19$\pm$0.32) & $UVW1$=16.87$\pm$0.05\\
00033818035 & 7209.362 & 0.078$\pm$0.010 (0.91$\pm$0.24) & $U$=16.27$\pm$0.03\\
00033818036 & 7209.959 & 0.082$\pm$0.010 (0.64$\pm$0.16) & $UVW2$=17.83$\pm$0.06\\
00033818038 & 7211.251 & 0.072$\pm$0.012 (0.64$\pm$0.23) & $UVM2$=20.47$\pm$0.95\\
00033818039 & 7211.956 & 0.063$\pm$0.010 (0.60$\pm$0.19) & $UVW1$=16.83$\pm$0.05\\
00033818040 & 7212.489 & 0.087$\pm$0.011 (0.79$\pm$0.20) & $UVW1$=16.78$\pm$0.05\\
00033818041 & 7212.876 & 0.055$\pm$0.009 (1.01$\pm$0.31) & $U$=16.25$\pm$0.03\\
00033818042 & 7213.210 & 0.090$\pm$0.011 (0.72$\pm$0.18) & $U$=16.29$\pm$0.03\\
00033818044 & 7214.291 & 0.079$\pm$0.008 (0.60$\pm$0.12) & $UVW2$=17.81$\pm$0.10\\
00032767003 & 7284.945 & 0.066$\pm$0.003 (0.72$\pm$0.07) & $U$=16.11$\pm$0.04\\
00034282008 & 7524.973 & 0.075$\pm$0.011 (0.54$\pm$0.16) & \\
\hline
\end{tabular}
\end{table*}
\setcounter{table}{0}
\begin{table*}
\centering
\footnotesize
\caption{Continued}
\setlength{\tabcolsep}{3.3pt}
\begin{tabular}{lccc}
\hline\hline
ObsID & HJD-2\,450\,000. & Count Rates ($HR=H/S$) & UVOT \\
      &                  & (cts\,s$^{-1}$)         & (mag) \\
  \hline
00034282009 & 7536.200 & 0.112$\pm$0.015 (0.71$\pm$0.19) & \\
00034282010 & 7540.523 & 0.088$\pm$0.014 (0.61$\pm$0.20) & \\
00034282011 & 7550.759 & 0.097$\pm$0.026 (0.69$\pm$0.37) & \\
00034282013 & 7572.083 & 0.014$\pm$0.013 (1.00$\pm$1.42) & \\
00034282014 & 7575.484 & 0.073$\pm$0.021 (0.17$\pm$0.13) & \\
00034282015 & 7576.331 & 0.041$\pm$0.030 (0.53$\pm$0.66) & \\
00034282016 & 7580.965 & 0.063$\pm$0.010 (1.04$\pm$0.32) & $U$=15.64$\pm$0.04\\
00034282018 & 7592.261 & 0.066$\pm$0.006 (0.70$\pm$0.13) & \\
00034282019 & 7606.088 & 0.043$\pm$0.015 (1.09$\pm$0.70) & \\
00034282022 & 7619.060 & 0.069$\pm$0.010 (0.89$\pm$0.26) & \\
00034282025 & 7640.273 & 0.092$\pm$0.012 (0.96$\pm$0.26) & \\
00034282026 & 7645.064 & 0.068$\pm$0.017 (0.70$\pm$0.35) & \\
00034282030 & 7688.916 & 0.063$\pm$0.013 (0.79$\pm$0.32) & \\
00093148001 & 7849.560 & 0.091$\pm$0.013 (0.75$\pm$0.22) & \\
00034282068 & 7872.201 & 0.102$\pm$0.013 (0.88$\pm$0.22) & \\
00093146003 & 7879.516 & 0.089$\pm$0.014 (0.49$\pm$0.16) & \\
00034282072 & 7899.906 & 0.043$\pm$0.009 (0.98$\pm$0.43) & \\
00093146005 & 7907.787 & 0.060$\pm$0.006 (0.98$\pm$0.18) & \\
00093148003 & 7910.982 & 0.048$\pm$0.005 (0.80$\pm$0.17) & \\
00093146006 & 7921.373 & 0.063$\pm$0.015 (0.99$\pm$0.46) & \\
00093146007 & 7936.120 & 0.086$\pm$0.012 (0.77$\pm$0.21) & \\
00034282078 & 7942.377 & 0.086$\pm$0.012 (0.74$\pm$0.21) & \\
00034282079 & 7947.361 & 0.079$\pm$0.013 (0.59$\pm$0.19) & \\
00034282080 & 7947.992 & 0.061$\pm$0.008 (0.51$\pm$0.15) & \\
00034282081 & 7948.923 & 0.067$\pm$0.009 (0.79$\pm$0.21) & \\
00034282082 & 7949.618 & 0.078$\pm$0.013 (0.65$\pm$0.22) & \\
00093146008 & 7950.203 & 0.086$\pm$0.010 (0.84$\pm$0.19) & \\
00034282084 & 7951.740 & 0.089$\pm$0.018 (0.47$\pm$0.20) & \\
00034282086 & 7954.301 & 0.073$\pm$0.008 (0.67$\pm$0.14) & \\
00034282089 & 7956.693 & 0.088$\pm$0.008 (0.76$\pm$0.14) & \\
00093148005 & 7971.868 & 0.074$\pm$0.017 (0.73$\pm$0.35) & \\
00093148008 & 8017.971 & 0.067$\pm$0.011 (0.58$\pm$0.20) & \\
00010451001 & 8096.174 & 0.074$\pm$0.005 (0.82$\pm$0.11) & $UVW1$=16.78$\pm$0.05\\
00010451002 & 8105.767 & 0.045$\pm$0.004 (1.06$\pm$0.20) & $U$=16.07$\pm$0.04, $UVW2$=17.82$\pm$0.11\\
00010451003 & 8116.128 & 0.038$\pm$0.004 (1.02$\pm$0.20) & $UVW1$=16.70$\pm$0.04\\
00010451004 & 8126.220 & 0.056$\pm$0.004 (0.92$\pm$0.14) & $UVW2$=17.58$\pm$0.06\\
00010451005 & 8135.691 & 0.063$\pm$0.005 (0.87$\pm$0.13) & $UVW1$=16.67$\pm$0.06\\
00010451006 & 8145.784 & 0.063$\pm$0.004 (0.76$\pm$0.10) & $U$=16.06$\pm$0.04, $UVW2$=17.63$\pm$0.06\\
00010451007 & 8155.787 & 0.070$\pm$0.005 (1.09$\pm$0.15) & $UVW1$=16.84$\pm$0.05\\
00010451008 & 8166.616 & 0.062$\pm$0.005 (1.00$\pm$0.16) & $UVM2$=21.56$\pm$1.14, $UVW2$=17.58$\pm$0.07\\
00010451009 & 8176.201 & 0.074$\pm$0.005 (0.75$\pm$0.10) & $UVW1$=16.73$\pm$0.04\\
00010451010 & 8185.605 & 0.070$\pm$0.006 (0.72$\pm$0.12) & $UVW2$=17.78$\pm$0.17\\
00034282139 & 8242.850 & 0.068$\pm$0.039 (0.52$\pm$0.58) & \\
00094061005 & 8253.227 & 0.087$\pm$0.015 (0.99$\pm$0.33) & \\
00094061006 & 8266.706 & 0.044$\pm$0.007 (0.76$\pm$0.23) & \\
00094061008 & 8294.862 & 0.119$\pm$0.016 (0.63$\pm$0.17) & \\
00088806001 & 8356.197 & 0.064$\pm$0.007 (0.83$\pm$0.17) & $UVW1$=16.68$\pm$0.05\\
00088807001 & 8358.463 & 0.060$\pm$0.007 (0.69$\pm$0.15) & $UVW2$=17.53$\pm$0.06\\
\hline
\multicolumn{4}{l}{\ch}\\
2572  & 2486.975 & \\
16659 & 7038.160 & \\
\hline
\multicolumn{4}{l}{{\it Suzaku}}\\
402030010 & 4453.880 & \\
\hline
\end{tabular}
\\
\tablefoot{$^a$: $V=11.74\pm0.03,B=14.88\pm0.03, U=14.86\pm0.04, UVW1=16.75\pm0.07, UVM2=21.23\pm1.18,$ and $UVW2=17.67\pm0.08$.\\ For \sw, missing UVOT photometry corresponds to observations where \duz\ is out of the field of view, except for 00033818008 (UVM2 filter) in which the star remains undetected.\\The {\it ASCA} dataset mentioned in Sect. 2.2.4 has ObsID=20003000 and was taken on $HJD$=2\,449\,106.959. }
\end{table*}

\begin{figure*}
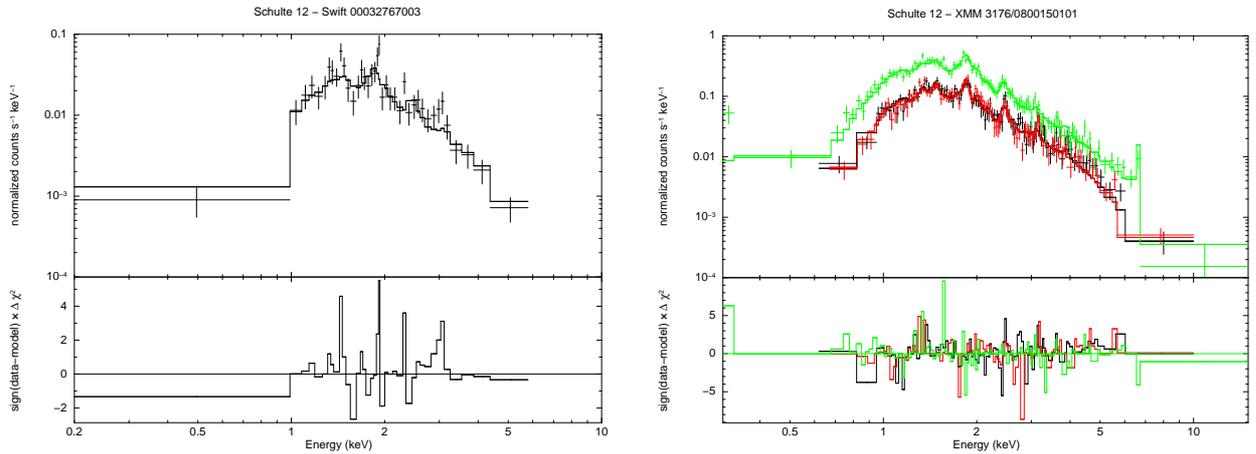

  \begin{center}
\includegraphics[width=6cm,angle=-90]{sw00032767003_src12.ps}
\includegraphics[width=6cm,angle=-90]{3176_cygob212.ps}
  \end{center}
\caption{X-ray spectrum of \duz\ (Left: \sw\ data of observation 00032767003, Right: \xmm\ data taken on rev. 3176 with pn in green, MOS1 in black, and MOS2 in red) along with their best-fit models (see Table \ref{fits}) and the residuals.}
\label{specfig}
\end{figure*}

\begin{figure*}
  \begin{center}
    \includegraphics[width=4.5cm,bb=34 180 560 605, clip]{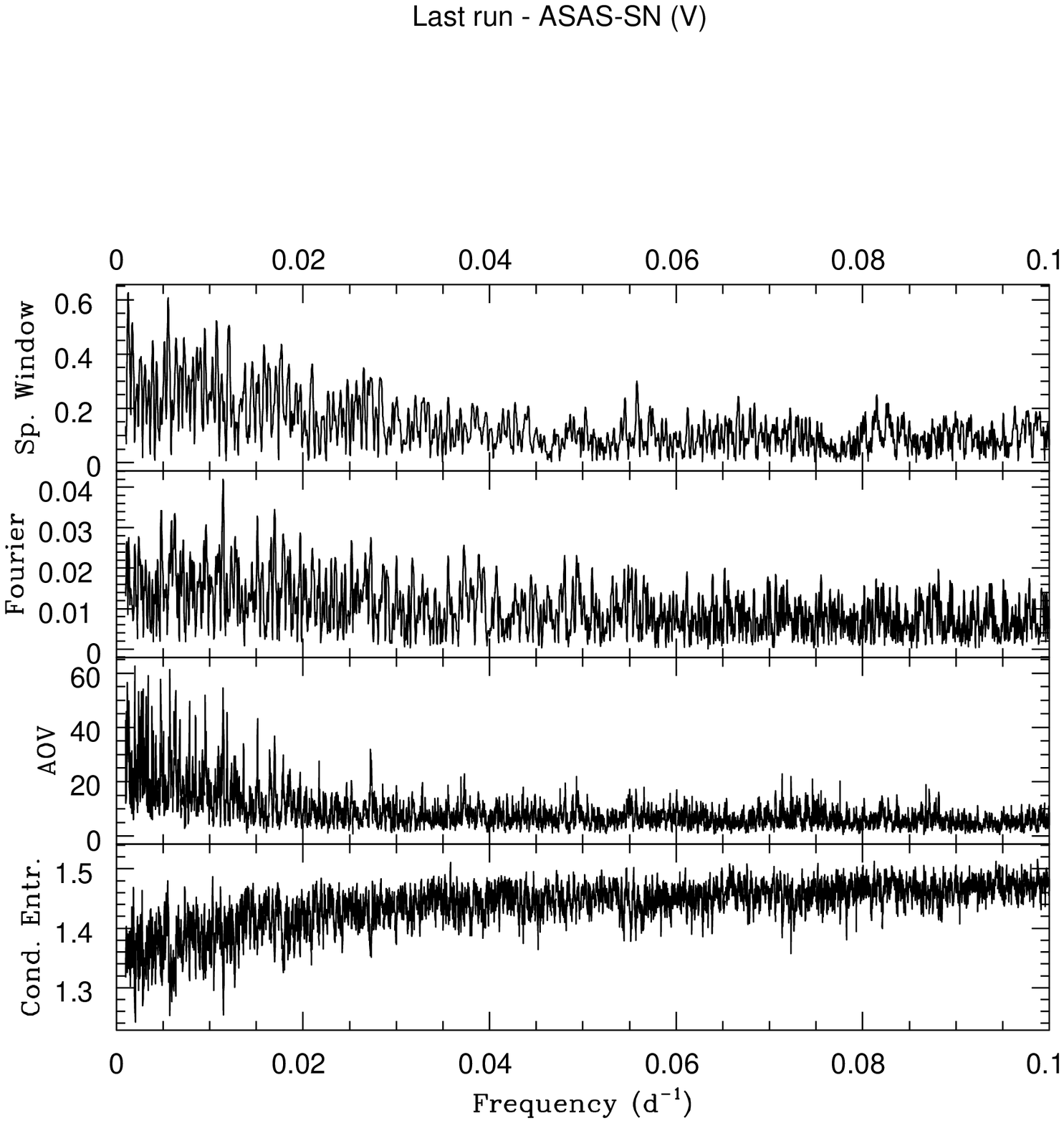}
    \includegraphics[width=4.5cm,bb=34 180 560 605, clip]{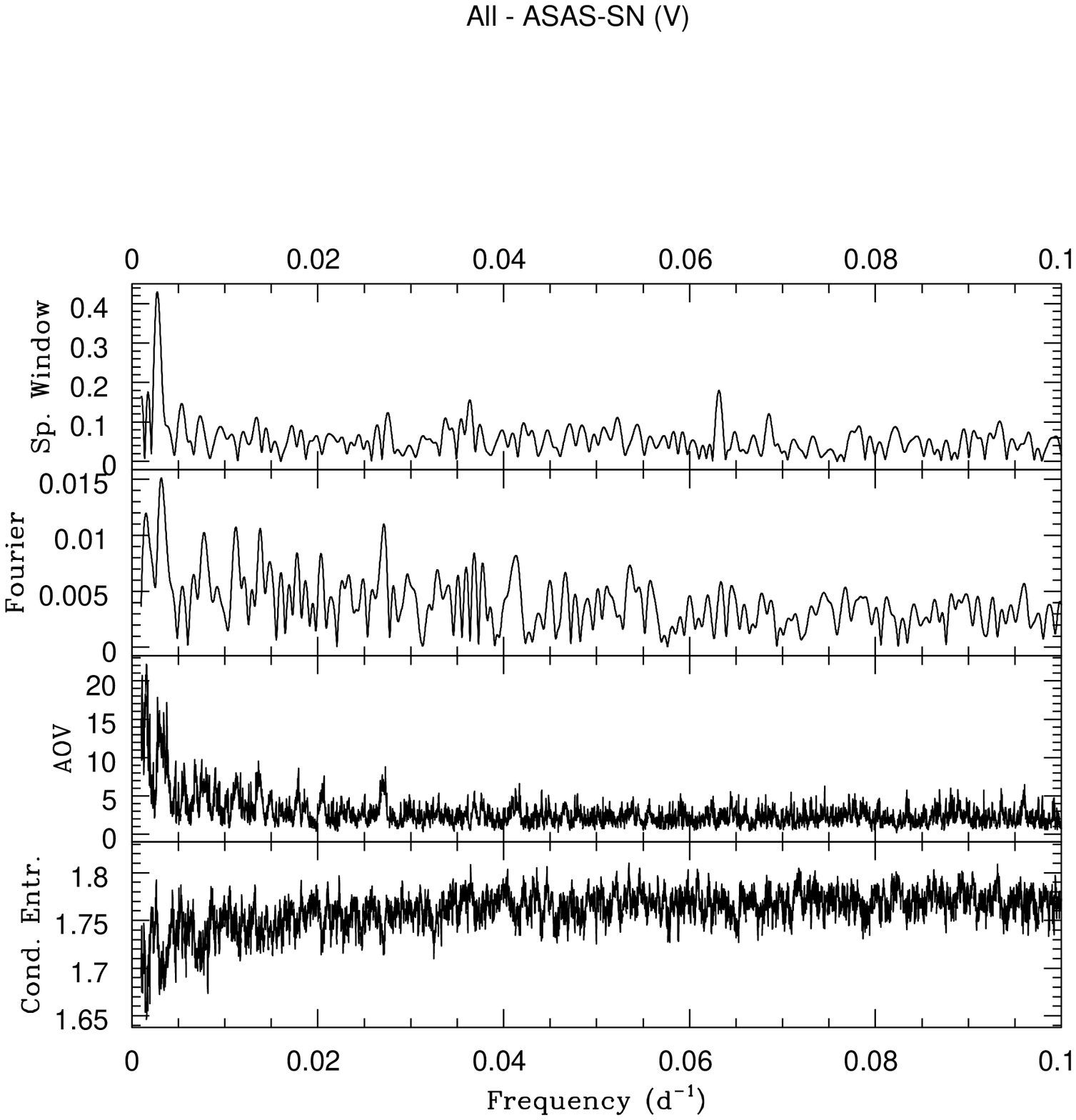}
    \includegraphics[width=4.5cm,bb=34 180 560 605, clip]{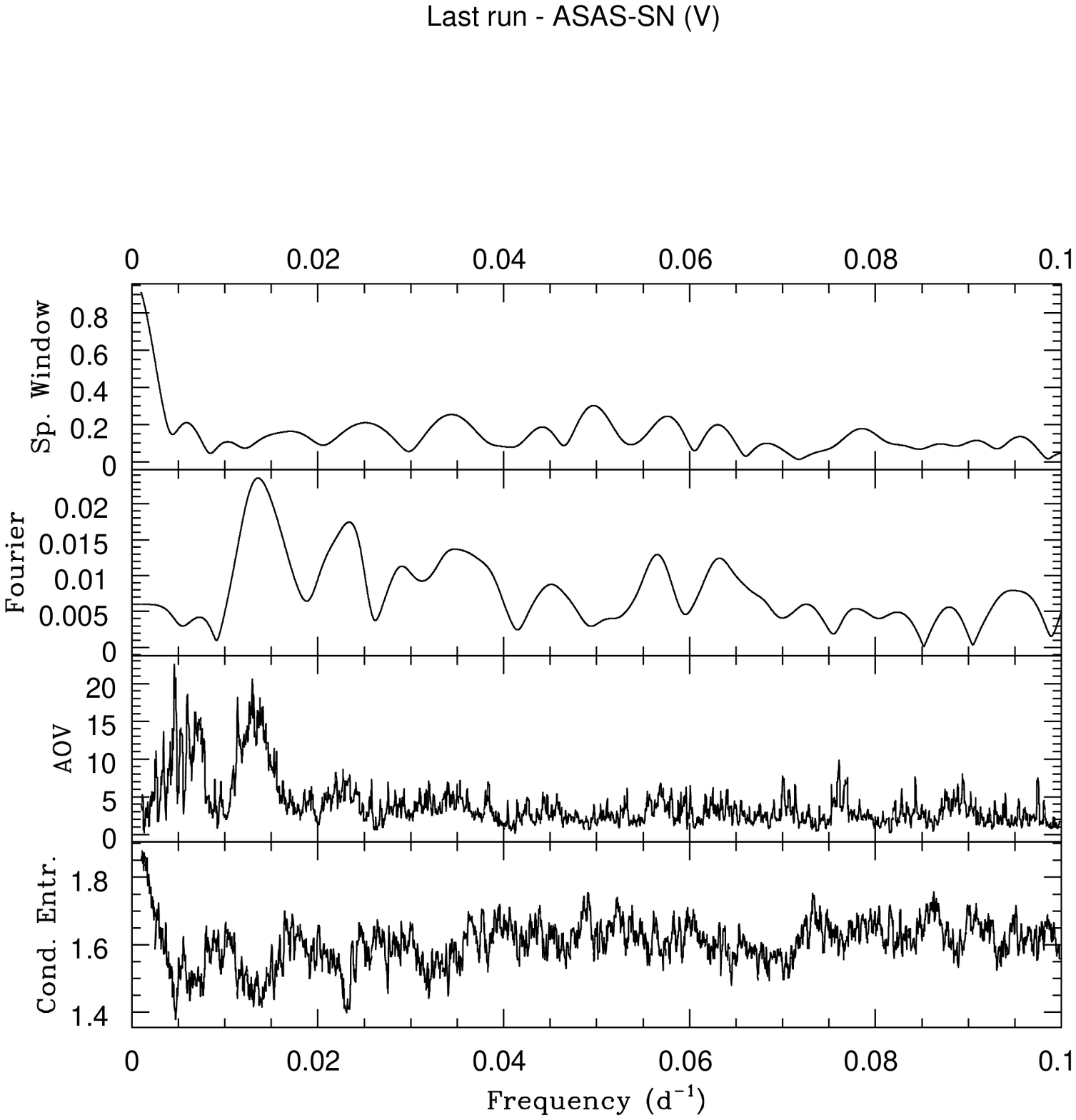}
    \includegraphics[width=4.5cm,bb=34 180 560 605, clip]{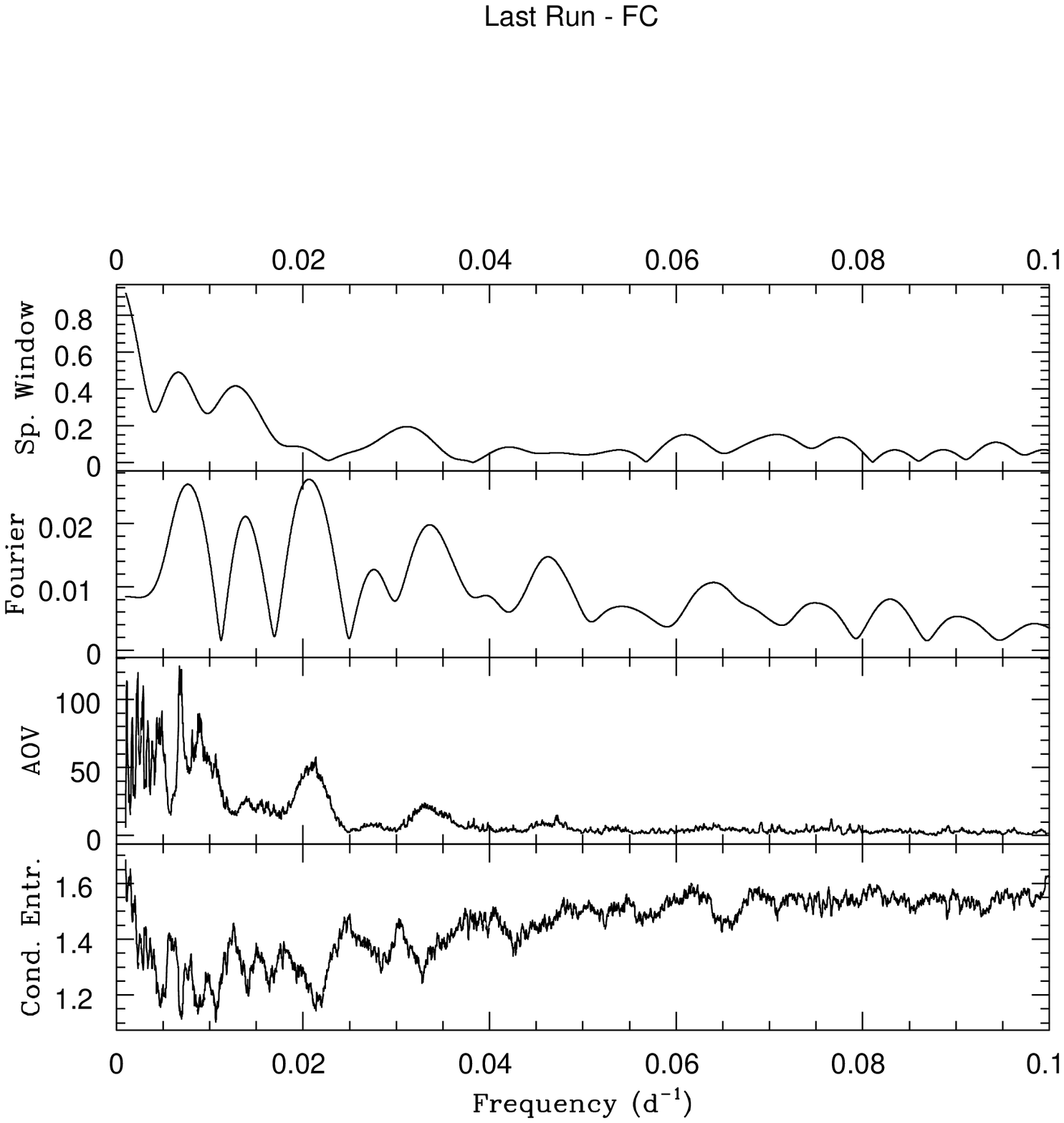}
  \end{center}
\caption{Results from period searches on the visible light curves. From left to right: Results for OMC data, ASAS-SN data (all runs and last run only), and our own photometry. In each panel, from top to bottom: Periodogram resulting from the Fourier method (with its spectral window), the AOV method, and the conditional entropy method (see text for details).}
\label{addps}
\end{figure*}

\begin{figure}
  \begin{center}
\includegraphics[width=8cm]{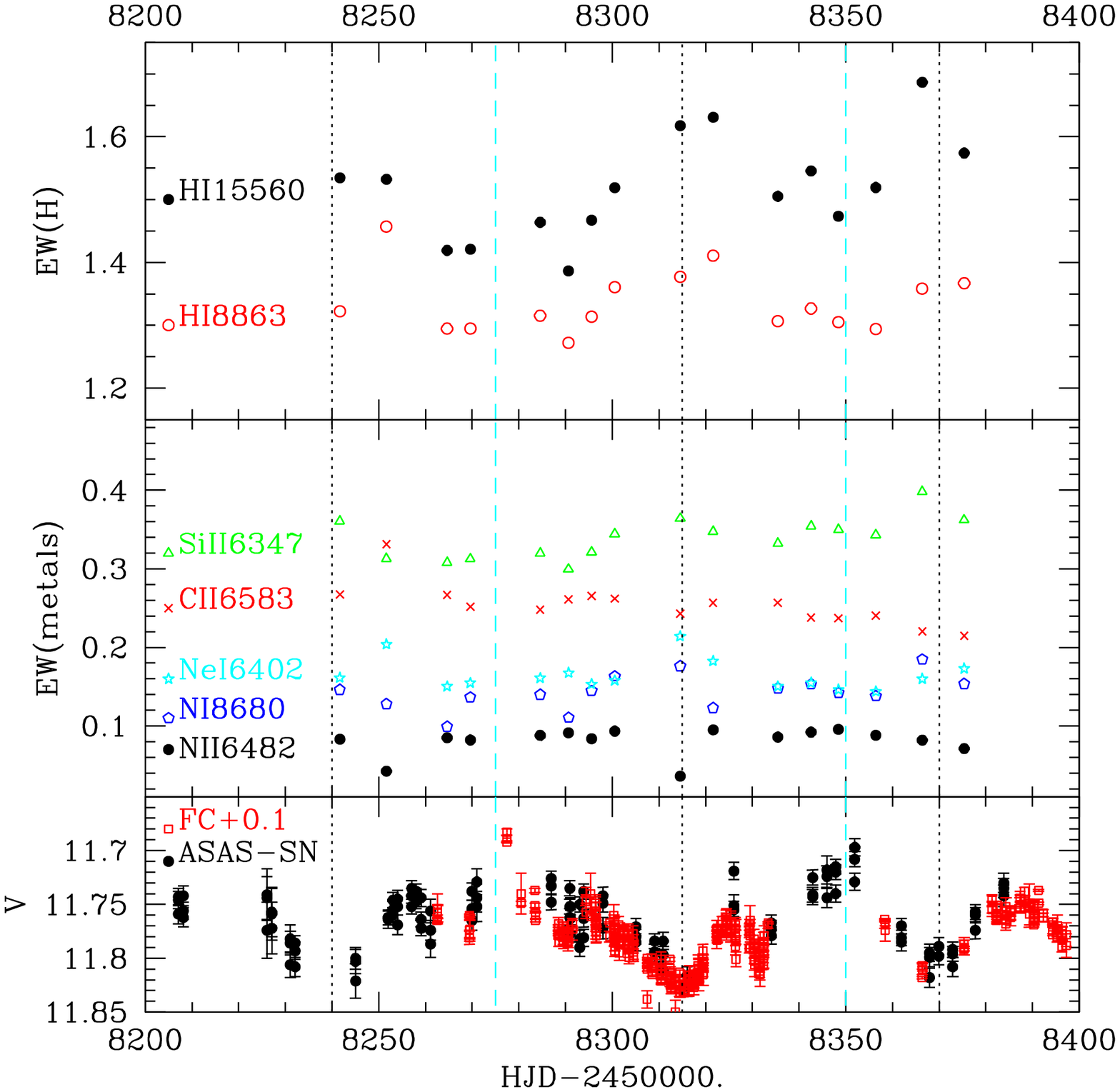}
  \end{center}
\caption{Variation of the line strengths and the photometry in 2018. Formal errors on EWs are similar to the symbol size. The dashed cyan and dotted black vertical lines indicate times of maximum and minimum brightness, respectively.}
\label{sptype}
\end{figure}

\end{appendix}

\begin{acknowledgements}
Y.N. and G.R. acknowledge support from the Fonds National de la Recherche Scientifique (Belgium), the Communaut\'e Fran\c caise de Belgique, the European Space Agency (ESA) and the Belgian Federal Science Policy Office (BELSPO) in the framework of the PRODEX Programme (contract XMaS). S.C. acknowledges support through DFG projects SCH 1382/2-1 and SCHM 1032/66-1. We thank the \sw\ team for their kind assistance. This project has received funding from the European Union's Horizon 2020 research and innovation programme via the OPTICON project under grant agreement No 730890. This material reflects only the authors views and the Commission is not liable for any use that may be made of the information contained therein. We are grateful to the Calar Alto service mode observers (M.\ Moreno, B.\ Arroyo, A.\ Guijarro, I.\ Hermelo, R.P.\ Hedrosa, P.\ Martin, G. Bergond \& A. Fern\'andez) for their help in collecting the Carmenes data. Some results were obtained with the Mercator telescope, operated by the Flemish Community on the island of La Palma at the Spain Observatory del Roche de los Muchachos of the Instituto de Astrofisica de Canarias. Mercator and Hermes are supported by the Funds for Scientific Research of Flanders (FWO), the Research Council of KU Leuven, the Fonds National de Recherche Scientifique (FNRS), the Royal Observatory of Belgium, the Observatoire de Gen\`eve, and the Th\"uringer Landessternwarte Tautenburg. L. M. thanks all the observers who collected the data with Hermes. ADS and CDS were used for preparing this document. 
\end{acknowledgements}

\end{document}